\documentstyle[preprint,aps,eqsecnum]{revtex}
\begin{document}
\draft
\title
{A NONPERTURBATIVE CALCULATION OF BASIC  \\
 CHIRAL QCD PARAMETERS WITHIN DYNAMICAL  \\
 EQUATIONS APPROACH TO QCD AT LOW ENERGIES}
 
\author{V. Gogohia, Gy. Kluge and M. Priszny\'ak}
 
\address{ RMKI, Department of Theoretical Physics,
Central Research Institute for Physics, \\
H-1525,  Budapest 114,  P. O. B.  49,  Hungary}
 
\maketitle
 
\begin{abstract}
Basic chiral QCD parameters (the pion decay constant, the quark
and gluon
condensates, the dynamically generated quark mass, etc) as well as
the vacuum energy density have been calculated from first
principles within a recently proposed dynamical equations approach
to QCD at low energies. The zero modes enhancement (ZME) model of
quark confinement and dynamical chiral symmetry breaking
(DCSB) based on the solution to the Schwinger-Dyson
(SD) equation for the quark propagator in the infrared (IR) domain
was used for this purpose.
There are only two independent
quantities by means of which  calculations should be done
within our approach. Our unique input data was chosen to
be the pion decay constant in the chiral limit given by the chiral
perturbation theory at the hadronic level (CHPTh).
Phenomenological estimates of these quantities, as well as the
vacuum energy density, are in good
agreement with our numerical results.
The nonperturbative vacuum structure which emerges
from the ZME model, appears to be well suited to describe
quark confinement, DCSB, current-effective (dynamical)-constituent, as
well as constituent-valence quark transformations, the
Okubo-Zweig-Iizuka (OZI) rule, dimensional transmutation, etc. The
importance of the instanton-type fluctuations in the true QCD vacuum
for the ZME model is also emphasized. This allows to predict new,
more
realistic values for the vacuum energy density (apart from the
sign, by definition, the bag constant) and the gluon condensate.
\end{abstract}


\vfill
\eject

\section{ Introduction}

  Why is it so important to calculate properties of observed
physical states such as hadrons from first principles? An
answer is almost obvious: because that reproduces the dynamical
structure and the symmetry properties of the underlying
fundamental theory - QCD [1] with
minimal assumptions and in a model independent way. At
the same time this is a very difficult task. First of all, it is
a problem of large distances (infrared (IR) or low energy region).
In contrast to short distances, little is known about QCD at
these distances. The IR region is responsible for
the nonperturbative dynamics, which plays a vital role in the
formation of the above mentioned physical states. In order to
correctly describe nonperturbative dynamics, it is necessary to
bring two important nonperturbative ideas in QCD together: quark
confinement [2] and dynamical (or equivalently spontaneous)
chiral symmetry breaking (DCSB) [3-8]. There is a close
connection between these nonperturbative effects (see, for
example, Refs. 9-16). At high energies QCD is under the control of
the phenomenon of asymptotic freedom [17] which justifies the
use of the perturbation theory in this limit. At low energies,
however, QCD is governed by
$SU_L(N_f) \times SU_R(N_f)$ chiral symmetry
($N_f$ is the number of different flavors) and its dynamical
breakdown in the vacuum to the corresponding vectorial subgroup
[18]. Thus to understand chiral limit physics means to correctly
understand the dynamical structure of low energy QCD. A realistic
calculation of various physical quantities in this limit becomes
important. The extrapolation to the chiral
limit is also important in lattice calculations [19].
DCSB and the chiral limit itself are fine
tuning effects in QCD and they need nonperturbative methods for
their analytical and numerical investigations.
 
   Recently we have proposed a nonperturbative and microscopic
approach to QCD at low energies [16, 20], the dynamical
equations approach to QCD. It consists of two parts: the chiral
perturbation theory at the quark level (CHPTq) and the dynamical
(or nonperturbative) quark propagator approach to
QCD at large distances (low energies). After the removal of
the possible singularities in a self-consistent way [20]
(see also the next section), the CHPTq represents the
elements of the $S$-matrix, reproducing the corresponding physical
quantities, in terms of formal Taylor expansions, in powers of
small external momentum (the momentum of a physical state, for
example, pion). In principle a general scheme of this method is
rather simple. It is worth drawing the
quark diagram (or diagrams) of the corresponding elements
of the $S$-matrix (see Fig. 2).
Then the analytic expressions of these diagrams
(written down in accordance with the Feynman rules for the full
quark propagators and proper vertices) or the corresponding elements
of the $S$-matrix themselves should be expanded in
powers of the corresponding external momentum (if any).
In other words, the CHPTq yields correct
analytic expressions of the amplitudes for physical quantities
in a systematic way: term by term in power series expansions of
the external (physical) momentum. On the other hand, this makes
also possible to factorize dependence on the physical (hadron) and
unphysical (quark) momenta which, obviously, greatly simplifies
further analysis. Certainly the convergence of these
series is assumed. At this stage
we have no ideas about convergence yet since that depends strongly
on the solutions to the quark Schwinger-Dyson (SD) equations.
 
The coefficients of these expansions (which are functions of an
unphysical momentum only because of the above mentioned
factorization) supply the solutions to the corresponding SD
equations for the quark propagator.
So the CHPTq admits incorporation of the quark
confinement phenomenon, in a self-consistent way, through
realistic, nonperturbative and confinement-type solutions to the
above mentioned quark SD equations. These solutions, which
correctly describe especially the IR region in QCD,
are required for a reliable calculation of various important
low-energy physical QCD parameters. Some of these solutions
have already been published [12 ,13, 16]. They were obtained
on the basis of the zero modes enhancement (ZME) effect, a
consequence of the
(possible) nonperturbative IR divergences in QCD (see
Section 3). Owing to the CHPTq and dynamical quark propagator
approaches, which complement each other, it is possible to
analytically
investigate, and calculate numerically, low energy QCD physics
from first principles in a self-consistent way.
 
 To perform realistic calculations of basic chiral QCD parameters
is our main purpose.
We confirm numerically our expectation that the dominant
contributions to the values of all chiral QCD parameters come
from large distances, while the contributions from short
and intermediate distances can only be treated as small
perturbative corrections. In this context let us note that
nonperturbative analytic calculations (advocated here) of the
properties of the physical states do not contradict lattice
calculations which also provide the opportunity
to compute them from first principles. Moreover, analytic and
lattice [21], and especially lattice SD equations [22],
calculations should complement each other in order to get correct
physical information for the description of the properties of the
physical states.
 
  The paper is organized as follows. In Section 2 a CHPTq solution
of the flavour non-singlet, chiral axial-vector Ward-Takahashi
(WT) identity in the framework of DCSB is briefly discussed.
Section 3 is devoted to a summary of our dynamical quark
propagator approach to QCD at large distances in the ZME model
of quark confinement and DCSB. In Section 4, we derive explicit
expressions for the quark and gluon contributions to the vacuum
energy density. In Section 5, within
our approach, we derive expressions for the pion decay constant in
the current algebra (CA) representation, quark condensate,
dynamical quark mass, etc, suitable for numerical calculations. We
present our numerical results in Section 6. Section 7 is devoted
to our main conclusions. A detailed discussion of a possible
dynamical mechanism of quark confinement, DCSB and
problems related to the
ZME effect in the QCD vacuum is given in Section 8 as well as a
possible
dynamical explanation of the famous Okubo-Zweig-Iizuka (OZI)
selection rule. The need of the instanton-type fluctuations for
our model of the QCD vacuum is also qualitatively discussed in
this section. Some related
models and important problems are discussed in Sections 9 and 10,
respectively.

 \section{ CHPT at the quark level solution of the chiral,
           axial-vector WT identity}

For the sake of the reader's convenience and for some other
purpose as well, let us briefly describe the CHPTq approach
here, applying it to the flavour non-singlet, chiral
axial-vector WT identity in the framework of DCSB
[20]. As it is well known, the Goldstone states
associated with the dynamically broken chiral symmetry
should be considered as quark-antiquark bound-states [5-8, 20].
The wave functions of these bound-states can be restored
from the axial-vector, chiral WT identity as the coefficient
function at the pole $q^2=0$ of the corresponding axial-vector
vertex, according to DCSB.
Moreover, the correct treatment of this identity within the CHPTq
provides  complete information on its regular
piece at zero momentum transfer $(q=0)$ (see below). The main
problem involved in the
analytical calculations of the bound-states amplitudes and axial
vertices through the corresponding axial-vector WT identities
is their dependence on arbitrary form
factors. These form factors are usually neglected by
formally assuming that they are of order $g^2$ in the
coupling constant [5, 9, 23, 24]. It has been shown in [20]
that these perturbative arguments are not valid
for treating such a nonperturbative effect like DCSB is.
 
     After these general remarks, let us consider now the flavour
non-singlet,
axial-vector WT identity in the chiral limit $m_0 = 0$, where $m_0$
denotes the current ("bare") mass of a single light quark.
\begin{equation}
   iq_\mu\Gamma^i_{5\mu}(p+q,p)=\left({\lambda^i \over
 2}\right)
   \{\gamma_5S^{-1}(p)+S^{-1}(p+q)\gamma_5 \},
\end{equation}
where $q=p'-p$ is the momentum transfer (the external momentum)
and the quark propagator is given by
\begin{equation}
         -iS(p)= \hat pA(-p^2)+B(-p^2).
\end{equation}
 In connection with (2.1), one has to point out that, in
general, $\lambda^i$ is a $SU(N_f)$ flavour matrix  and, in the
massless case, the quark propagator is proportional to the unit
matrix in the flavour space.
The inverse of the quark propagator of Eq. (2.2) is expressed as
\begin{equation}
      \{-iS(p)\}^{-1} = \hat p \overline A(-p^2)-\overline
B(-p^2),
\end{equation}
where
\begin{eqnarray}
    \overline A(-p^2) &=& A(-p^2)D^{-1}(-p^2), \nonumber\\
          \overline B(-p^2) &=& B(-p^2)D^{-1}(-p^2), \\
           D(-p^2) &=& p^2 A^2(-p^2)-B^2(-p^2) .\nonumber
\end{eqnarray}
 
 Using the standard decomposition of the inverse quark
propagator Eq. (2.3), DCSB at the fundamental quark level
can be implemented by the following condition
\begin{equation}
\bigl\{ S^{-1}(p),  \gamma_5 \bigr \}_+ = i \gamma_5
2 \overline B(-p^2) \ne 0,
\end{equation}
so that the $\gamma_5$ invariance of the quark propagator is
broken
and the measure of this breakdown is the double of the dynamically
generated quark mass function  $2 \overline B (-p^2)$.
This condition leads to the zero mass
boson (Goldstone state) in the flavour, axial-vector WT identity
(2.1). Indeed, it follows from (2.1) that one gets a nonzero
dynamical quark mass, defined by (2.5), if and only
if $\Gamma^i_{5\mu}(p+q,p)$ has a pseudoscalar pole
at $q^2=0$ (a dynamical singularity which determines Goldstone
state) and vice versa [4, 5, 20].
 
     The Lorentz invariance and parity allow one to write
the axial-vector vertex in terms of twelve independent form
factors [20, 25] as
\begin{eqnarray}
\Gamma^i_{5\mu}(p+q,p) = &\bigl( {\lambda^i\over 2}\bigr) \gamma_5
\{ \gamma_\mu G_1 + p_\mu G_2 + p_\mu \hat p G_3
  + \hat p \gamma_\mu G_4 + q_\mu G_5 + \hat q q_\mu G_6
  + \hat q p_\mu G_7 \hfill \nonumber\\
       &+ \hat q\gamma_\mu G_8
 + \hat p q_\mu G_9 + \hat p \hat q p_\mu G_{10} +
 \hat p \hat q q_\mu G_{11} +
 \hat p \hat q \gamma_\mu G_{12} \},
 \end{eqnarray}
where $(j=1,2,3...12)$.
\begin{equation}
G_j \equiv G_j(-p^2,-{p'}^2,-q^2) = G_j(-p^2,-(p+q)^2,-q^2)
\equiv G_j(-(p+q)^2),
\end{equation}
In order to avoid unphysical (kinematic)
singularities at $(p'q)=0$ in what follows, the external
momentum $q$ (transfer momentum) and
the initial momentum of the quark $p$ are chosen as
independent momenta in the decomposition
(2.6) (thus, $(pq)=0$ only at $q=0$,
see Ref. 26). We will be especially interested in the
behaviour of the form factors (2.7) as
functions of the external momentum $q$ (momentum of a massless
pion). So all the above mentioned physical singularities will be
singularities of these form factors at $q=0$. Only
the form factors $G_5,G_6,G_9,$ and $G_{11}$ can contribute to
the dynamical pole-like structure at  $q^2=0$ of the
axial-vector vertex (2.6) because only these ones are multiplied
by $q_\mu$,
whereas the other ones can also be singular at
$q=0$ (see below). Dependence on
$p^2$ at this stage remains, in principle,  arbitrary and
will be determined, within the CHPTq approach,
by nonperturbative solutions of the corresponding
quark SD equations.
Observe that one should find a
nonperturbative quark propagator which is regular at the zero
point and approaches
the free propagator at infinity (asymptotic freedom [17]).
 
   Substituting the decomposition (2.6) into the WT
identity (2.1), one obtains
four important relations between the form factors $G_j$, namely
\begin{eqnarray}
&&G_1 + q^2 G_6 + (pq) G_7 = \overline A(-(p+q)^2)   \\
&&(p q) G_2 + q^2 G_5 + q^2 G_8 = \overline B(-(p+q)^2) +
                                 \overline B (-p^2)   \\
&&(p q) G_3 + q^2 G_9 +q^2 G_{12}= \overline A(-(p+q)^2) -
\overline A (-p^2)   \\
&&G_4 + q^2 G_{11} + (p q) G_{10} = 0 ,
\end{eqnarray}
where $\overline A(-(p+q)^2)$ and
$\overline B(-(p+q)^2)$ are defined by (2.4)
with the substitution $p \rightarrow p+q$.
 
     In order to calculate certain important physical parameters,
e.g. the pion decay constant, the meson form
factors, their charge radii, etc., one needs to know
the bound-state BS meson
wave function, restored from the WT identity (2.1) as a residue at
pole in the axial-vector vertex (2.6) and its regular piece
at small momentum transfer
$q$ and especially at zero momentum transfer $q=0$ (see below).
For this reason it is necessary to transform the
singular part of the axial-vector vertex in (2.6) into its regular
counterpart, thus making the corresponding form factors free of
singularities at $q=0$. This can be done in a complete analogy
with the Ball and Chiu method [26] to remove
unphysical (kinematic) singularities from the corresponding
vertex. For the sake of simplicity, in what follows, let us call
this the "Ball and Chiu procedure".
 
  Indeed, from the exact relations (2.8-2.11) it follows that the
form factors $G_1$ and $G_4$ are regular functions of their
arguments from the very beginning. The form factors $G_j$
(j=2,3,7,10) have singularities of the type $(pq)^{-1}$
at $q=0$ and the form factors $G_8$ and $G_{12}$ may have
singularities of the type $q^{-2}$ at $q=0$, which may
be mixed up with the dynamical singularity at $q^2=0$.
However one can explicitly show that these form
factors, as well as the form factors $G_1$ and $G_4$, are regular
functions of $q$ at small $q$ from the very beginning [20].
As mentioned above, only the form factors $G_5, G_6, G_9$ and
$G_{11}$ need to have dynamical pole singularities at $q^2=0$,
whereas the form factors $G_8$ and $G_{12}$ can have also the same
singularities at this point. For this reason,
let us express these form factors  as follows
(j=5,6,9,11 and also j=8,12)
\begin{equation}
G_j(p,q) = {1 \over {q^2}} R_j(p,p) + G_j^R(p,q),
\end{equation}
where  $R_j(p,p)$  and $G_j^R (p,q)$ are the residues and regular
parts of the corresponding form factors.
 
 In the framework of CHPTq, let us now write down the regular form
factors $G_1$ and $G_4$
in terms of formal Taylor series in powers of
small external momentum $q$ as follows $(j=1,4)$
\begin{equation}
G_j(-(p+q)^2) = \sum^{\infty}_{n=0}{h^n \over n!}
G^{(n)}_j(-p^2)
\end{equation}
where
\begin{equation}
      h = - 2(pq)-q^2.
\end{equation}
The same expansions for the quark invariant functions are in order
\begin{equation}
\overline A(-(p+q)^2) = \sum^{\infty}_{n=0}{h^n \over n!}
\overline A^{(n)}(-p^2)
\end{equation}
and
\begin{equation}
\overline B(-(p+q)^2) = \sum^{\infty}_{n=0}{h^n \over n!}
\overline B^{(n)}(-p^2)
\end{equation}
respectively. Here and in what follows, as well as
in expansions (2.13-2.14), differentiation is understood
with respect to the argumentum $(-p^2)$ and, for
example, $\overline B^{(0)}(-p^2) \equiv \overline B(-p^2)$.
 
  By completing the Ball and Chiu procedure (for details
see our paper [20]), one is now able to decompose the
initial axial-vector
vertex (2.6) into pole (dynamical) and regular parts as follows
\begin{equation}
\Gamma^i_{5\mu}(p+q,p) =
- {q_\mu \over {q^2}} F_{\pi} G_5^i(p+q,p)
+ \Gamma^{iR}_{5\mu}(p+q,p),\\
\end{equation}
where the BS bound-state amplitude is
\begin{equation}
G_5^i(p+q,p)= - {1\over{F_{\pi}}}\left({\lambda^i \over
2}\right)\gamma_5 G(p+q,p),
\end{equation}
with
\begin{equation}
G(p+q,p) = 2 \overline B(-p^2)+\hat q R_6(-p^2)+\hat p \hat q R_{11}(-p^2)
\end{equation}
and the arbitrary form factors are the residues of the
corresponding form factors (2.12).
  The regular part now is determined as follows
\begin{equation}
\Gamma^{iR}_{5\mu}(p+q,p) = \bigl( {\lambda^i\over 2}\bigr)
\gamma_5 \{ \gamma_\mu G_1 + p_\mu G_2 + p_\mu \hat p G_3
+ \hat p \gamma_\mu G_4 + O_{\mu}(q) \}
\end{equation}
where $O_{\mu}(q)$ defines the terms of order $q$ and they play
no further role. At zero momentum transfer, the
regular part also depends on the same form factors
$R_6$ and $R_{11}$ as in (2.19) and is given by
\begin{eqnarray}
G_1(-p^2) &=& \overline A (-p^2) - R_6(-p^2)  \nonumber\\
G_2(-p^2) &=& - 2 \overline B'(-p^2) \nonumber\\
G_3(-p^2) &=& - 2 \overline A'(-p^2) \nonumber\\
G_4(-p^2) &=& - R_{11}(-p^2).
\end{eqnarray}
The system (2.21) is nothing else but the conditions for
cancellation of the dynamical singularities at $q=0$ for the
corresponding form factors [20]. Dependence on
the same arbitrary form factors in (2.19) and (2.21) indicates
that the self-consistent separation of the singular (pole)
part from the regular one in this vertex is not yet finished.
Indeed, it is easy to see that the terms multiplied by these
arbitrary form factors also give rise to nonvanishing
contributions to
the axial-vector vertex when $q$ goes to zero. There is no
explicit reason for keeping these terms in the pole part of the
corresponding axial-vector vertex in (2.17). Moreover, in order to
completely untangle the pole and regular parts, they
also should
be transmitted from the pole part to the regular part of the
vertex (transmission effect).
Finally one obtains
\begin{equation}
\Gamma^i_{5\mu}(p+q,p) = \left( {\lambda^i \over 2}
\right) \gamma_5
{q_\mu \over {q^2}} [ 2 \overline B(-p^2)]
+ \Gamma^{iR}_{5\mu}(p+q,p),\\
\end{equation}
where now
\begin{eqnarray}
\Gamma^{iR}_{5\mu}(p+q,p) = \bigl( {\lambda^i\over 2}\bigr)
\gamma_5 \{[\gamma_\mu G_1+{q_\mu \hat q \over {q^2}}R_6(-p^2)]
+p_\mu G_2 + p_\mu \hat p G_3  \hfill \nonumber\\
+[ \hat p \gamma_\mu G_4
+ {q_\mu \hat p \hat q \over {q^2}} R_{11}(-p^2)] +
O_{\mu}(q) \}
\end{eqnarray}
 
  On account of (2.21) it is easy to check that the sum in the
first brackets in (2.23) becomes simply
$\gamma_{\mu} \overline A(-p^2)$, i.e. an exact cancellation of
the unknown arbitrary form factor $R_6(-p^2)$ occurs.
In the same way, owing to (2.21), the contribution
from the second brackets completely disappears too, i.e. the
exact cancellation of the arbitrary form factor
$R_{11}(-p^2)$ takes place as well.
Thus the axial-vector vertex
$\Gamma^i_{5\mu}(p+q,p)$ in (2.17) has, indeed, a dynamical
Goldstone pole singularity at $q^2=0$, explicitly shown in (2.22),
which corresponds to the massless pion with residue
\begin{equation}
G(p,p) = R_5(-p^2) = 2 \overline B(-p^2)
\end{equation}
proportional to the pion decay constant  $F_\pi$ (2.18)
[5, 8, 9, 20, 23, 24]. The regular part of the vertex
at zero momentum transfer ($q=0$) becomes
\begin{equation}
\Gamma^{iR}_{5\mu}(p,p) = \left( {{\lambda^i} \over
 2}\right)\gamma_5 \{\gamma_{\mu}
\widetilde G_1+p_{\mu}G_2+p_{\mu}\hat p G_3 +\hat p
\gamma_{\mu} \widetilde G_4 \} ,
\end{equation}
where, evidently, all form factors are functions of $(-p^2)$
only and expressed as follows
\begin{eqnarray}
\widetilde G_1(-p^2) &=& \overline A (-p^2)  \nonumber\\
G_2(-p^2) &=& - 2 \overline B'(-p^2) \nonumber\\
G_3(-p^2) &=& - 2 \overline A'(-p^2) \nonumber\\
\widetilde G_4(-p^2) &=& 0.
\end{eqnarray}
Here the primes denote
differentiation with respect to the argumentum $(-p^2)$.
This is the general CHPTq solution to the regular piece
of the axial-vector vertex at zero momentum transfer (2.25)
in the chiral limit and it does not depend on the arbitrary form
factors at all.
 
  Expressions (2.22-2.26) indicate that the
self-consistent separation of dynamical pole singularity from the
regular part in the corresponding axial-vector vertex now is
completed. We call attention to that the arbitrary form
factors $R_6(-p^2)$ and $R_{11}(-p^2)$ can not be put "by hand"
to zero in (2.21) in order to obtain (2.26). An exact
cancellation, because of the above mentioned transmission effect,
takes place between the same (arbitrary) form factors entering the
bound-state amplitude up to terms of order $q$ (2.18-2.19) and
the regular part of the axial-vector vertex at zero
momentum transfer $q=0$ (2.20-2.21). These form factors become
necessarily the same in order to cancel dynamical singularities at
$q=0$ in the axial-vector vertex [20]. Only this allows
to completely untangle the pole and the regular parts in it.
 
    The regular part at zero momentum transfer is now
determined involving neither arbitrary form factors nor
the BS bound-state amplitude which coincides with the
residue at the pole. The form factor $G_9$ could, but
did not, contribute to the pole-like structure of the vertex
from the very beginning. The form factors
$G_6$ and $G_{11}$, as shown above, give rise to the same
(finite) contributions to the pole, as well as to the regular
parts of the vertex, so they cancel each other at zero momentum
transfer
(transmission effect). The form factor $G_5$ alone determines
the pole-like structure at $q^2=0$ of the axial-vector vertex.
Its residue coincides with the BS amplitude at zero momentum
transfer, Eq. (2.24). This result is well-known [5, 8, 9,
20, 23, 24], and we
only recast it within the CHPTq approach in terms of our form
factors
in the general decomposition of (2.6-2.7). A new exact result
appears
in Eqs. (2.25-2.26), which makes possible to express the regular
part of the axial-vector vertex at zero momentum transfer in terms
of the dynamical
quark propagator variables alone. As mentioned above, these form
factors ($G_2, G_3$ and $G_4$) had previously been neglected by
assuming that they were of order $g^2$ in the coupling constant
[9, 23, 24]. This solution
(2.25-2.26) leads to the new nonperturbative expression for the
pion decay constant in the Jackiw-Johnson (JJ)
representation [4, 20].
Needless to say, there is no hope for an exact solution
of the BS-type integral equation for the regular part of the
axial-vector vertex even at zero external momentum transfer.
 
  Note that the quantities of physical
interest (bound-state amplitudes and regular parts) can be
restored from the corresponding axial WT identity up to terms
of $O(q^0)$ only. This is in agreement with the general situation
when the WT identity can provide nontrivial information only on
quantities at zero momentum transfer (longitudinal terms) and the
terms of order $q$ and higher (transverse terms) always remain
undetermined. In other words, the correct treatment of this
identity provides exact information on the first terms of the
corresponding Taylor series (in powers of the external momentum)
for the bound-state amplitudes and the
regular parts of the corresponding vertices, etc. In many cases
of physical interest this information is completely sufficient for
calculating important physical parameters from low-energy physics
in a self-consistent way. In order to find terms of order $q$ and
higher in the CHPTq expansion of the BS bound-state amplitude
(2.18-2.19), it
is necessary to develop the CHPTq in the framework of the
corresponding
BS integral equation (which is now in progress) or investigate the
corresponding BS integral equation in some other
approximation as it was done, for example, in Ref. 27.
 
 Incidentally, let us note that from (2.1) and
(2.17-2.22), in general, it follows
\begin{equation}
iq_\mu\Gamma^{iR}_{5\mu}(p+q,p) = \left({\lambda^i \over
 2}\right) \{\gamma_5S^{-1}(p)+S^{-1}(p+q)\gamma_5 \} +
   iF_{\pi} G^i_5 (p+q,p),
\end{equation}
so differentiation with respect to $q_{\nu}$, on account of Eq.
(2.19), and then setting $q=0$ gives
\begin{equation}
i \Gamma^{iR}_{5\mu}(p,p) = \left({\lambda^i \over
 2}\right) \partial_{\mu} S^{-1}(p) \gamma_5 +
i \Delta^i_{5\mu}(p,p),
\end{equation}
where
\begin{equation}
 \Delta^i_{5\mu}(p,p) = \left({\lambda^i \over
 2}\right) [ \gamma_{\mu} R_6(-p^2) - \hat p \gamma_{\mu}
R_{11}(-p^2) ] \gamma_5.
\end{equation}
Thus, in order to find these arbitrary form factors, it is
necessary to solve the BS integral equation for the regular part
(2.28) at zero momentum transfer ($q=0$) in some appropriate
approximation
for the BS scattering kernel or to solve the BS integral equation
up
to terms of order $q$ for the pion bound-state amplitude (2.18-2.19)
\begin{equation}
G_5^i(p+q,p)= - {1\over{F_{\pi}}}\left({\lambda^i \over
2}\right)\gamma_5
[2 \overline B(-p^2)+\hat q R_6(-p^2)+\hat p \hat q R_{11}(-p^2)]
\end{equation}
in the same approximation. The above mentioned CHPTq within the BS
formalism is the most appropriate tool for this purpose.
It makes sense also to assume
some very similar nonperturbative ansatz for the arbitrary form
factors $R_6(-p^2)$ and $R_{11}(-p^2)$ (see Subsection A in
Section 5 below).
 
  Concluding this section a few remarks are in order.
In our paper [20] it has been already explained why
the above mentioned CHPTq formal Taylor expansions, in powers of
small external momentum $q$, are valid when the quark momentum
$p$ is also small. Introduction of the large mass scale parameter
$\mu$ is necessary in the CHPTq approach to make sense of the
formal Taylor expansions at small $p$. So, even in the domain of
small $p$, contributions from small external momentum $q$ in
(2.13-2.16)
are supressed by a factor of $\Lambda_{\chi} /\mu \ll 1$, where
$\Lambda_{\chi}$ is the scale of DCSB at the hadronic level - the
scale of the effective field theory (CHPTh) [28].
It is worth noting that if one formally identifies
the large mass scale parameter $\mu$ with the mass of the heavy
quark $m_Q$ then the same factor of suppression appears as in the
heavy quark expansion [29].

\section{A dynamical quark propagator approach to QCD
         at low energies}
 
  As it was already pointed out in the introduction, the IR or low
energy region (large distances) is responsible for
the nonperturbative
effects in QCD. The most important ones of these effects are quark
confinement and DCSB.
 In order to study these and other interesting effects
one should develop a nonperturbative approach to QCD at
large distances. The system of the SD
equations with the corresponding Slavnov-Taylor (ST) identities
for Green's functions [1, 30] can serve as an adequate and
effective tool for a nonperturbative approach to QCD (see also
recent reviews in Refs. 31). One must keep in mind
that an infinite system of nonlinear integral
SD equations, complemented by the corresponding ST identities
contains, in principle, the full dynamical information of
quantum field theory. One of the major and long-standing problems
in quantum
field theory is to find a gauge-invariant approximation scheme
to the SD equations and the corresponding ST identities since one
holds no hope for an exact solution. It seems to us that our
approach
to QCD at large distances [12, 13, 16] provides a reasonable
solution to this essential problem though in the quark sector
only.
 
 The central
role in our approach belongs to the quark propagator. A
correct description and the numerical calculation of the
properties of the physical particles (e.g. pions)
is extremely difficult because one needs to know the
quark propagator that  satisfies some necessary conditions.
A quark propagator must not be an explicitly gauge-dependent
quantity. It should be nonperturbative and preferably
regular at the zero point; it should have no poles
(confinement-type solution), and it should correspond
to the dynamical breakdown of chiral symmetry (see below).
A quark propagator must also be free of ghost complications,
provided it was obtained in any covariant gauge and, finally, it
must have a correct
asymptotic behaviour at infinity, i.e. it  must asymptotically
approach the free propagator at infinity (asymptotic freedom).
 
Making only one
widely accepted dynamical assumption that  the full gluon
propagator becomes IR singular like (hereafter referred to as the
enhancement of the zero modes in the QCD true vacuum (see the next
subsection))
\begin{equation}
D_{\mu\nu}(q) \sim (q^2)^{-2} ,
\qquad q^2 \rightarrow 0
\end{equation}
in the covariant gauge, such a quark propagator
has recently been found [12, 13, 16].
There exist direct and indirect arguments in favour of this
asymptotically IR sigular behaviour of the full gluon propagator
(3.1). It makes
sense to remind the reader only of a few of them. These have
nothing
to do with the potential concept relevant for the constituent
quark model (CQM) [32] despite its phenomenological success [33]
for single-hadron states.
 
  I. In the above mentioned case the Wilson loop defined as
\begin{equation}
W(c) = \langle Tr P_c \exp ig \oint A_{\mu} dx_{\mu} \rangle =
Tr \exp \left\{- g^2 \oint dx_{\mu} dy_{\nu} D_{\mu\nu}(x-y) +...
\right\},
\end{equation}
obeys an area law, indicative of quark confinement.
Indeed, if the full gluon propagator is singular like (3.1) as
the momentum goes to zero, then
$D_{\mu\nu}(x-y) \sim \ln (x - y)$ and
\begin{equation}
W(c) \sim \exp [- \sigma A(c)],
\end{equation}
where $A(c)$ is the minimal area of some surface bounded by curve
$c$. This is the Wilson criterion for quark confinement [2, 34].
For the free gluon propagator,
$D_{\mu\nu}(q) \sim (q^2)^{-1}$, then $D_{\mu\nu}(q) \sim
(x - y)^{-2}$ and the Wilson loop behaves as follows
\begin{equation}
W(c) \sim \exp [- \mu P(c)],
\end{equation}
where $P(c)$ is the perimeter of curve $c$. Therefore no statement
about confinement can be deduced.
 
II. The cluster property of the Wightman functions in QCD fails
and this allows such a singular behaviour like (3.1)
for the full gluon propagator in the IR domain [35].
 
III. The form factor of the full gluon
propagator (see, (3.6)-(3.7) below) is nothing else but the
running coupling constant. Then the ansatz (3.1) leads to the
strong coupling behaviour of the Callan-Symanzik-Gell-Mann-Low
(CS-GML) function
$\beta(g)$ at large distances ( $q^2 \rightarrow 0$). In this
case, the CS-GML function is always negative, i.e. there is no IR
stable fixed point at all. This behaviour of the
strong coupling regime ($g \rightarrow \infty$) is usually
refered to as IR slavery, indicative of confinement [1, 36].
 
IV. After the pioneering works of Baker, Ball and Zachariasen
in the axial gauge and Mandelstam in the covariant (Landau)
gauge [37], the consistency of the singular asymptotics (3.1) with
the direct solution of the full gluon propagator in the IR domain
was repeatedly confirmed (see, for example, Refs. 38, our paper
[16] and references therein).
Despite many problems and inconsistencies due to the complicated
mathematical structure and the highly nonlinear nature of the SD
equation for the full gluon propagator in QCD, the deep IR
singular
asymptotics (3.1) should be considered as well-established.

\subsection{The zero modes enhancement model of quark confinement
            and DCSB}

    Today there are no doubts that the dynamical mechanisms of
quark confinement and DCSB are closely related to the complicated
topological structure of the QCD nonperturbative vacuum [1, 15,
39] but, unfortunately, a detailed dynamical picture of these
nonperturbative effects is not yet known.  Also it becomes clear
that the nonperturbative IR divergenses, are closely related, on
one
hand, to the above mentioned nontrivial vacuum structure, on the
other hand, they are important as far as the large scale
behaviour of QCD is concerned [1, 37, 39]. If it is true that QCD
is an IR unstable theory (has no IR stable fixed point) then the
low-frequency modes of the Yang-Mills fields should
be enhanced due to the nonperturbative
IR divergences. So the gluon propagator diverges faster than
$(q^2)^{-1}$ at small $q$, in accordance with (3.1) - the
zero modes enhancement (ZME) effect in QCD.
If, indeed the low-frequency components of the virtual fields
in the true vacuum have a larger amplitude than those of the bare
(perturbative) vacuum [37], then the Green function  for a single
quark should be reconstructed on the basis of this effect.
It is important that the possible effect of the ZME (3.1) is our
primary dynamical assumption. We will consider this effect as
a very similar confining ansatz for the full gluon propagator in
order to use it as input information for the quark SD equation.
 
In what follows, for
the convenience of the reader, we present the above
mentioned reconstruction in its most general important features,
leaving additional details for our paper [16].
  Let us begin from some general remarks of mathematical nature.
Such a singular behaviour of the full gluon propagator requires
the
introduction of a small IR regulation parameter $\epsilon$
in order to define  the initial SD equations and ST
identities in the IR region (postponing the
precise definition of the distribution $(q^2)^{-2}$ in
$n$-dimensional Euclidean space). Because of this,  the
quark propagator and other Green's functions become dependent,  in
general, on this IR regulation parameter $\epsilon$,  which
is to be set to zero at the end of the computation
($\epsilon \rightarrow 0^+$). For the sake of brevity, this
dependence will be always understood but it will not be indicated
explicitly.
 
 There exist only two different types of
behaviour of the quark propagator with respect to $\epsilon$ in
the $\epsilon \rightarrow 0^+$ limit.
 If the quark propagator does not depend
on the $\epsilon$ - parameter in the $\epsilon \rightarrow 0^+$
 limit then one obtains the IR
finite (from the very beginning) quark propagator. In this
case quark confinement is understood as the disappearance of the
quark propagator pole on the real axis at the point $p^2 = m^2$,
where $m$ is the quark mass. Such an interpretation of quark
confinement comes, apparently, from Preparata's massive quark
model (MQM) [40] in which quarks were approximated by
entire functions. A quark propagator may or may not
be an entire function, but in any case the first order pole
disappears (see references in our paper [16]). On the other
hand, a quark
propagator can vanish after the removal ($\epsilon \rightarrow
0^+$) of the IR regulation
parameter $\epsilon$. A  vanishing quark propagator is also a
direct manifestation of quark confinement.
Apparently, this understanding of quark confinement follows
from two-dimensional QCD with $N_c$ large limit [41].
 
   Closing these general remarks, let us consider the exact,
unrenormalized SD equation for the quark
propagator in momentum space (Fig. 1)
\begin{equation}
S^{-1}(p) = S^{-1}_0(p) + g^2 C_F \int {d^nq\over {(2\pi)^n}}
\Gamma_\mu(p, q) S(p-q)\gamma_\nu D_{\mu\nu}(q),
\end{equation}
where  $C_F$  is the eigenvalue of the quadratic Casimir
operator in the fundamental representation. Other notions are
obvious.
From now on we will suppose that the IR region is
effectively decoupled from the UV region, following the paper of
Pagels [36]. The standard UV renormalization program should be
performed after completion of our program.
 
The full gluon propagator in an arbitrary covariant gauge is
\begin{equation}
D_{\mu\nu}(q) = - i \left\{ \left[ g_{\mu\nu} -
{{q_\mu q_\nu}\over {q^2}} \right]
{1\over {q^2}} d(-q^2,  a) + a {{q_\mu q_\nu}\over
{q^4}} \right\},
\end{equation}
where $a$  is a gauge fixing parameter ($a = 0,$  Landau
gauge).
 
  Assuming that in the IR region
\begin{equation}
d(-q^2,  a) = \left( {{\mu^2}\over {-q^2}} \right) +
\beta(a) + O(q^2),
\qquad q^2 \rightarrow 0,
\end{equation}
where $\mu$ is the appropriate  mass scale parameter,  we obtain
the above mentioned generally accepted form of the IR singular
asymptotics for the full gluon propagator (3.1) (enhancement of
the zero modes). In accordance
with (3.7), hereafter we keep only the leading
and next-to-leading terms of the corresponding Green's functions
expansions in the IR region under control, omitting always the
terms of order $q^2$.
 
  In connection with (3.6) and (3.7), let us note that
despite the approximation of the full gluon propagator (3.6) with
its deep IR singular asymptotics (3.7), formally in the whole
range $[0,\infty)$, such kind of behaviour may occur
only in the nonperturbative (IR) region.
This means that correct solutions to the quark
SD equations should manifest the existence of a scale at which
nonperturbative effects become essential. As we will show
explicitly below, our solutions to the quark SD equation really
contain such a characteristic scale intrinsically.
 
    In order to actually define an initial SD equation
(3.5) in the IR region (at small momenta) let us
apply the gauge-invariant dimensional regularization
method of 't Hooft and Veltman [42] in the limit $n = 4 + 2
\epsilon,\quad \epsilon \rightarrow 0^+$.
 Here and below $\epsilon$ is the above mentioned
small IR regulation parameter. Now we
consider the SD equations and the corresponding quark-gluon ST
identity in Euclidean space
($d^nq \rightarrow i d^nq_E,  \quad q^2  \rightarrow - q^2_E,
\quad p^2  \rightarrow -p^2_E,  \quad $
but for simplicity the Euclidean subscript  will be omitted).
 
 Let us use, in the sense of distribution theory, the relation
[43]
\begin{equation}
(q^2)^{-2+\epsilon} = {{\pi^2}\over {\epsilon}} \delta^4(q) +
(q^2)^{-2}_+ + O(\epsilon), \qquad \epsilon \rightarrow 0^+,
\end{equation}
which implies  that the full gluon propagator (3.6) in the
IR region behaves like
\begin{equation}
D_{\mu\nu}(q) = \epsilon^{-1} \bar D_{\mu\nu}(q),
\qquad \epsilon \rightarrow 0^+,
\end{equation}
in the  $\epsilon \rightarrow 0^+ $ limit
and $\bar D_{\mu\nu}(q) $  exists as $\epsilon  \rightarrow 0^+$.
Here and
below $(q^2)^{-2}_+$ is the functional acting on the
main (test) functions according to the so-called "plus
prescription" standard  formulae [43] (see below).
 
  The singularity (3.8-3.9) is the only initial singularity
in our
approach to QCD at large distances. All other Green's functions
will be considered as regular functions of their arguments.
In the distribution theory their singular dependence leads to
more
complicated SD equations for the IR finite (renormalized) quark
propagator and therefore requires special treatment. With these
caveats let us proceed to the realization of the renormalization
program in order to evaluate nonperturbative IR divergences,
provided by the strongly singular confining ansatz (3.8-3.9).
 
   Substituting (3.8-3.9) into the SD equation (3.5),  we
obtain the quark propagator expansion in the IR
region (in four-dimensional Euclidean space)
\begin{eqnarray}
S^{-1}(p) = S^{-1}_0(p)&+&{1\over {\epsilon}}{\tilde g}^2
\Gamma_\mu(p, 0) S(p)\gamma_\mu  \nonumber \\
&+& c_1 \int {d^4q\over {(2\pi)^4}}
\Gamma_\mu(p, q) S(p-q)\gamma_\nu t_{\mu\nu}(q) (q^2)^{-2}_+
\nonumber\\
&+& c_2 \int {d^4q\over {(2\pi)^4}}
\Gamma_\mu(p, q) S(p-q)\gamma_\nu T_{\mu\nu}(q,a) (q^2)^{-1}
\nonumber\\
&+&O(\epsilon), \quad  \epsilon \rightarrow 0^+,
\end{eqnarray}
where ${\tilde g}^2 = C_F {3\over 4} g^2 {\mu}^2 {\pi}^2
(2\pi)^{-4}$ and
\begin{equation}
t_{\mu\nu}(q) = [g_{\mu\nu} - {q_{\mu} q_{\nu} \over q^2}]
\end{equation}
is the transverse tensor and the tensor
\begin{equation}
T_{\mu\nu}(q,a)
= \beta(a) t_{\mu\nu}(q) + a{q_\mu q_\nu \over q^2}
\end{equation}
explicitly depends on a gauge-fixing parameter $a$.
 
Within the Gelfand's and Shilov's distribution theory [43] one
has
\begin{eqnarray}
\int {d^4q\over {(2\pi)^4}}
\Gamma_\mu(p, q) S(p-q)\gamma_\nu t_{\mu\nu}(q) (q^2)^{-2}_+
\nonumber\\
= \int {d^4q\over {(2\pi)^4}} t_{\mu\nu}(q) \left\{
\Gamma_\mu(p, q) S(p-q) - \Gamma_\mu(p,0) S(p) \right\}
\gamma_\nu (q^2)^{-2}.
\end{eqnarray}
So, expanding in powers of $q$ and keeping the terms of order
$q^{-2}$ ( the Coulomb order terms), in agreement with (3.7), from
(3.10) and on account of (3.13), one finally obtains
\begin{eqnarray}
S^{-1}(p) &=& S^{-1}_0(p) + {1\over {\epsilon}}{\tilde g}^2
\Gamma_\mu(p, 0) S(p)\gamma_\mu  \nonumber \\
&+& c_1 \int {d^4q\over {(2\pi)^4}} t_{\mu\nu}(q) (q^2)^{-2}
\left\{- q^{\alpha} q^{\lambda}
\Gamma^{\alpha}_{\mu}(p,0) S^{\lambda}(p)
+{1 \over 2}q^{\alpha} q^{\beta} \Gamma^{\alpha\beta}_{\mu}(p,0)
S(p) \right\}\gamma_\nu  \nonumber \\
&+& \int {d^4q\over {(2\pi)^4}} \left\{ {c_1 \over 2}
t_{\mu\nu}(q) (q^2)^{-2} q^{\lambda} q^{\tau} \Gamma_\mu(p,0)
S^{\lambda\tau}(p) + c_2 T_{\mu\nu}(q,a) (q^2)^{-1}
\Gamma_\mu(p,0) S(p) \right\} \gamma_\nu
\nonumber\\
&+& O(\epsilon), \quad  \epsilon \rightarrow 0^+,
\end{eqnarray}
In derivation of this expansion, we have used
formal Taylor series for Green's functions as
\begin{equation}
\Gamma_{\mu}(p,q) = \Gamma_{\mu}(p,0) + \sum^{\infty}_{n=1}
{q^n \over n!} \left\{ {d^n \Gamma_{\mu}(p,q) \over dq^n}
\right\}_{q=0}
\end{equation}
and
\begin{equation}
S(p-q) = S(p) + \sum^{\infty}_{n=1}
{(-1)^n q^n \over n!} {d^n S(p) \over dp^n},
\end{equation}
where $q^n$ and $dq^n$ stand for
\begin{equation}
q^n = \prod^n_{m=1} q^{{\alpha}_m} \quad and \quad
dq^n = \prod^n_{m=1} dq^{{\alpha}_m},
\end{equation}
respectively. The same notation stands for $dp^n$ as
well. In (3.14), for example,
$\Gamma^{\alpha}_{\mu}(p,0)=\left\{ {d \Gamma_{\mu}(p,q) \over
dq^\alpha} \right\}_{q=0}$ and so on.
 
It is worth making a few remarks in advance. We will show,
by completing our renormalization program (see below), that the
finite terms (next-to-leading terms), coming from the
IR region and explicitly depending on a gauge fixing paratemer,
in any case, always become terms of order $\epsilon$. For
this reason they vanish as $\epsilon \rightarrow 0^+$,
like the finite term does (which is not written down) containing
the logarithm of the coupling constant and arising because of the
dimensional regularization method [42].
 
   As mentioned above, all Green's functions
become dependent generally on the IR regularization
parameter $\epsilon$.
In order to extract the finite Green's functions in
the IR region, we introduce the renormalized
(IR finite) quark-gluon vertex function at zero momentum
transfer and the quark propagator as follows
\begin{eqnarray}
\Gamma_\mu(p, 0) &=& Z_1(\epsilon) \bar \Gamma_\mu(p, 0),
\nonumber \\
S(p) &=& Z_2(\epsilon) \bar S(p),
\qquad \epsilon \rightarrow  0^+.
\end{eqnarray}
Here $Z_i(\epsilon)\ (i=1,2)$
are the corresponding IR renormalization constants.
The $\epsilon$-parameter dependence is indicated explicitly
to distinguish them from the usual UV renormalization
constants. In all relations, here and below, containing the
IR renormalization constants, the
$ \epsilon \rightarrow  0^+ $ limit is always assumed.
There are no restrictions on the
$ \epsilon \rightarrow  0^+ $ limit behaviour of the IR
renormalization constants, apart from the regular $\epsilon$
dependence of the IR renormalization constant $Z_2(\epsilon)$
of the quark wave function. This is required by the
quark confinement definition (mentioned above).
$\bar \Gamma_\mu(p, 0)$ and $\bar S(p)$ are the renormalized
(IR finite) Green's functions and they therefore do not depend
on $\epsilon$ in the
$ \epsilon \rightarrow  0^+ $ limit, i.e. they exist as
$ \epsilon \rightarrow  0^+ $.
 
  It is obvious that relations (3.18) and the analogous
relations for other Green's functions [16] are the most
general expressions, relating the
unrenormalized Green's functions to the renormalized (IR finite)
ones, if one assumes (in analogy with UV multiplicative
renormalizability (MR))
the property of MR of the IR nonperturbative divergences, arising
in the theory due to nonperturbative confining
ansatz (3.8-3.9) for the full gluon propagator in the IR
domain. This property was one of the main topics investigated
and proven in our paper [16].
 
  In connection with relations (3.8-3.9), Eq.(3.10) and
relations (3.18), let us make a few remarks in advance.
The correct treatment of such a strong singularity like (3.8-3.9)
within distribution theory [43] by the
gauge-invariant dimensional regularization method of t'Hooft and
Veltman [42] enabled us to extract the required class of test
functions
 in the renormalized quark SD equation. The test functions do
 consist of the quark propagator and the corresponding quark-gluon
vertex function, Eq.(3.10). By the renormalization program we
have
found the regular solutions for the quark propagator (see below).
For that very reason relation (3.8) is justified, it is
multiplied by the appropriate smooth test functions.
 
 From (3.14) and on account of (3.18) a cancellation of
nonperturbative IR divergences takes place if and only if (iff)
\begin{equation}
Z_1(\epsilon)Z_2^2(\epsilon) =  \epsilon Y_1, \qquad \epsilon
\rightarrow 0^+,
\end{equation}
with $Y_1$ being an arbitrary
finite constant. It is evident that this very condition and the
similar ones below, govern the concrete $\epsilon$ -dependence
of the IR renormalization constants which, in general,
remain arbitrary. From (3.19) it follows that the behaviour
of the vertex IR renormalization constant $Z_1(\epsilon)$
is determined completely  by the quark IR renormalization constant
$Z_2(\epsilon)$. Carrying out our renormalization program,
we show that all other IR renormalization constants
can also be expressed through the quark wave function
IR renormalization constant $Z_2(\epsilon)$ only.
 
  Expression (3.19) is a convergence condition for the quark
propagator. Because
of (3.19), the explicitly gauge-dependent terms (the
next-to-leading
terms) in the SD equation (3.14) become $\epsilon$ - order
 terms. For this reason these noninvariant terms vanish in the
$\epsilon \rightarrow  0^+ $ limit. From the formal Taylor
expansions (3.15) and (3.16) it follows that IR
renormalization properties of the lower (quark
propagator) and higher (quark-gluon vertex) Green's functions are
different. Indeed, if the derivatives of the quark propagator are
renormalized like the quark propagator itself then one
cannot say anything
about the derivatives of the quark-gluon vertex function since
we know its renormalization only at zero momentum transfer
(3.18). This is a principle difference between the nonperturbative
IR renormalization of any lower and higher Green's functions in
our model. At the same time it is easy to see that these terms
are always accompanied by proper powers of $q$, making the
terms containing, in general, the derivatives of the Green's
functions free from nonperturbative IR divergences. Obviously,
the finite next-to leading terms, coming from the IR region and
not containing the derivatives of the higher Green's functions
become, terms of order $\epsilon$. For this reason they
vanish in the $\epsilon \rightarrow  0^+$ limit, as well as the
above mentioned finite terms explicitly depending on a gauge
fixing parameter. Thus the leading and next-to-leading terms of
our
expansion for the quark propagator in the IR domain always become
not $explicitly$ dependent on a gauge-fixing parameter $a$. This
is
a general feature of our expansions in the IR region for various
Green's functions. The next-to-leading terms coming
from the IR region, which remain finite by completing our
renormalization program
(the ones containing the derivatives of higher Green's functions),
should be considered as a good approximation for the
intermediate region, which still remains $terra \ incognita$
in QCD. Apparently, the Gromes constant [33], which occurs
in the derivation of the quark-antiquark static potential
in the Wilson loop approach [44, 45], is the
nonrelativistic counterpart of these fully relativistic terms.
These terms (second line in Eq. (3.14)) are of the same Coulomb
order of magnitude ($ \sim q^{-2}$ in contrast to the
initial singularity (3.1)) as the corrections coming from the UV
region and they should be investigated, along with them,
elsewhere. The correct treatment of the enhancement
of the zero modes, given by (3.8) within the distribution
theory (3.12-3.16), nevertheless effectively converts the initial
strong IR singularity ($ \sim q^{-4}$) into a Coulomb-like
behaviour ($ \sim q^{-2}$) in the intermediate and UV regions,
which is
compatible with asymptotic freedom. Also, these terms are gauge
invariant in comparison with the terms displayed in the third line
of Eq. (3.16) which disappear after
completing the nonperturbative renormalization program.
 
  An important general observation is appropriate here. The ZME
model requires the full Green's functions, not the free ones, in
particular only the full quark-gluon vertex is relevant for our
approach. Also, from now on we will be interested only in
the leading terms of the corresponding Green's functions
expansions in the IR domain. The renormalization scheme to
remove all nonperturbative IR divergences, on a general ground,
depends only on these terms.
So one has (as $\epsilon \rightarrow  0^+ $ )
\begin{equation}
\bar S^{-1}(p) = Z_2(\epsilon) S^{-1}_0(p) + {\tilde g}^2Y_1
\bar \Gamma_\mu(p, 0) \bar S(p)\gamma_\mu,
\end{equation}
as the deep IR content of the quark SD equation in our model.
To conclude, let us stress once more that the mechanism
described above is
a general one, which shows clearly how to untangle the explicitly
gauge-independent and the explicitly gauge-dependent parts of the
IR piece of the quark propagator. The
next-to-leading terms which remain finite by completing the
nonperturbative IR renormalization program, because of the much
less
singular behaviour in the IR, are expected to produce only small
corrections to the contributions determined by the leading term.

\subsection{The ground solution}

  Absolutely in the same way should be reconstructed the ghost
self-energy and the corresponding ST identity for the quark-gluon
vertex [16]. We develop a method for the extraction of the
IR-finite Green's functions in QCD. The IR finiteness
of the Green's functions means that they do not depend on the
IR regulation parameter $\epsilon$ as $\epsilon \rightarrow 0^+$.
For this purpose, we have worked out a renormalization program
in order to cancel all the IR nonperturbative
divergences. It was possible to obtain a closed set
of the SD equations and the corresponding ST identity in the
quark sector.
By completing our renormalization program, we have explicitly
shown [16] that all Green's functions are IR MR.
In other words, multiplication by the quark IR renormalization
constant only makes possible to remove all nonperturbative
IR divergences from the theory.
  It is a general feature of our approach that the behaviour of
all Green's functions in the IR region depends only on the
quark wave function IR renormalization constant.
We also discover that our approach implies the existence of a
characteristic
scale at which confinement and other nonperturbative effects begin
to play a dominant role. If QCD confines then such a
characteristic scale must certainly exist.  We have shown
that for the covariant gauges the complications due to
ghost contributions can be considerable in our approach. We are
now able to formulate the main result of our approach
in terms of the following theorem [16].
 
${\underline{Quark \ confinement \ theorem}}$
 
Let us assume that the full gluon propagator in the covariant
gauge becomes infrared singular like $(q^2)^{-2}$ at small
momentum (enhacement of the zero modes). The quark SD equation has
then three and only three confinement-type solutions for the quark
propagator in the infrared region. Two of them are infrared
vanishing solutions after the removal of the infrared
regulation parameter. The third solution does not depend on the
infrared regulation parameter at all, but it
has no pole on the real axis in the complex momentum plane and
\underline{it implies dynamically chiral symmetry
breakdown (DCSB).}
 
  Basic chiral QCD parameters will be calculated
from the IR finite solution to the quark
SD equation. The closed set of equations in this case, obtained in
our paper [16], should read
 
\begin{equation}
S^{-1}(p)= S_0^{-1} (p) + \tilde g^2 \bar \Gamma_\mu(p,0)
S(p) \gamma_\mu,
\end{equation}
\begin{equation}
{1\over 2}\bar b(0) \bar  \Gamma_\mu(p,0) =
i \partial_{\mu} S^{-1}(p) - {1\over 2}\bar b(0) S(p)
 \bar \Gamma_\mu(p,0) S^{-1}(p).
\end{equation}
where
\begin{equation}
S^{-1}_0(p) = - i (\hat p - m_0),
\end{equation}
is the free quark propagator with $m_0$ being the current
("bare") mass of a single quark and $\tilde g^2$ includes the
mass scale parameter $\mu^2$, determining the validity of the
above mentioned deep IR singular asymptotic behaviour
of the full gluon propagator (3.7).
Let us recall also that
\begin{equation}
\partial_{\mu}S^{-1}(p) = {\partial \over
\partial p_{\mu}}S^{-1}(p)
\end{equation}
as well as that the IR finite quark renormalization constant
(which should multiply the free quark propagator in Eq.(3.21)) is
to be set to unity without losing generality (multiplicative
renormalizability) [16]. It is worth noting here that Eq.(3.21)
and Eq.(3.22) describe the leading terms of the corresponding
expansions of the quark SD equation and ST identity in the
IR region, respectively [16].
 
     In order to solve the system (3.21-3.22), it is
 convenient to represent the quark-gluon vertex
 function at zero momentum transfer as follows
\begin{equation}
\bar \Gamma_\mu(p,0) = F_1(p^2) \gamma_\mu + F_2(p^2)p_\mu
 + F_3(p^2)p_\mu \hat p + F_4(p^2) \hat p\gamma_\mu.
\end{equation}
Substituting this
 representation into the ST identity (3.22), one can express
 the scalar functions  $F_i(p^2) (i = 1,2,3,4 )$ in terms of the
 quark propagator scalar functions $A$ and $B$.
\begin{eqnarray}
 F_1(p^2)&=& - \overline A(p^2), \nonumber \\
 F_2(p^2)&=& - 2 \overline B'(p^2) - F_4(p^2),  \nonumber\\
 F_3(p^2)&=&  2 \overline A'(p^2),    \nonumber\\
 F_4(p^2)&=&  {{A^2(p^2)B^{-1}(p^2)} \over {E(p^2)}}.
\end{eqnarray}
 Here the prime denotes differentiation with respect to the
 Euclidean momentum variable  $p^2$ and
\begin{eqnarray}
\overline A(p^2) &=& A(p^2)E^{-1}(p^2), \nonumber\\
\overline B(p^2) &=& B(p^2)E^{-1}(p^2), \\
         E(p^2) &=& p^2 A^2(p^2) + B^2(p^2) .\nonumber
\end{eqnarray}
For the sake of convenience the ghost self-energy at
zero point $\bar b \equiv \bar b(0)$ will be included into
the definition of a new coupling constant (see below (3.31)).
 
   Proceeding now to the dimensionless variables
\begin{equation}
A(p^2) = \mu^{-2}A(t),\qquad B(p^2) = \mu^{-1}B(t),\qquad
 t=p^2/{\mu^2}
\end{equation}
 and doing some algebra, the initial system (3.21-3.22) can be
 rewritten as follows (normal form)
\begin{equation}
  A'= -2 \left\{ {1\over t} +
 {1\over \lambda} \right\}A
- {2\over \lambda}{1\over t} - {2\over \lambda}{1\over t} m_0 B,
\end{equation}
\begin{equation}
 B' = -{3\over 2} A^2 B^{-1} + {2\over \lambda} \left( m_0 A -
 B\right),
\end{equation}
 where $A \equiv A(t)$, \ $B \equiv B(t)$, and
here the prime denotes differentiation with
 respect to the Euclidean dimensionless momentum variable $t$.
 A new coupling constant is
\begin{equation}
\lambda = g^2[\overline b(0)]^{-1}(2\pi)^{-2},
\end{equation}
 where $\bar b(0)$  is the finite ghost self-energy at the
zero  point.
 
Thus our system (3.29-3.30) does not depend on the parameter
$\bar b(0)$ in explicit form. Instead of two
different parameters, i.e. initial coupling constant $g^2$ and
$\bar b(0)$, we
have only one parameter - the coupling constant $\lambda$
(3.31) and we therefore avoid the difficulties associated with
the unknown ghost contributions in the covariant gauge.
In (3.31) one can
recognize the effect of finite renormalization of the initial
coupling constant with the help of the ghost
self-energy, finite at zero point. This means that the ghosts,
having played their important role,
retire from the scene.
 
Let us note here that a dimensionless initial coupling
constant $g^2$, as well as $\lambda$ of (3.31), plays no
independent role in the presence of the initial
mass scale parameter $\mu$, which characterizes the
nonperturbative region (see the next section).
However, here we did not include it into $\mu$ because of two
reasons: 1) to retain the correspondence with contributions to the
quark SD equation, coming from the UV region, where this
possibility is lacking. 2) to explicitly demonstrate the
nonperturbative nature of our solutions.
 
   As it will be
shown later, this system always has a chiral symmetry
 breaking solution ($m_0 = 0,  B(t) \ne 0 $ ).
Our system (3.29-3.30)
 excludes the trivial solution $A=B=0$. Any nontrivial
 solution automatically breaks the $\gamma_5$ invariance of
the quark propagator  and it therefore leads to
 spontaneous chiral symmetry breakdown
($m_0 = 0, B(t) \ne 0$,  dynamical quark mass generation). In the
 general case ($m_0 \ne 0$), it is possible to show that
the  solutions of this system cannot have pole-like
 singularities at any finite point on the real axis
 in the whole complex $t$-plane [16]. This is a direct
 manifestation of quark confinement.
 
    Let us consider the chiral limit of this system, $m_0 = 0$. In
this case the initial  system (3.29-3.30)
can be exactly solved. The regular solutions are
\begin{equation}
A(t) = -\left({\lambda \over 2}\right)t^{-2}
\left\{\exp {\left(- {2\over \lambda}t\right)}-1 + {2\over
\lambda}t\right\}.
\end{equation}
and
\begin{equation}
B^2(t_0, t) = 3 \exp{(- {4\over \lambda}t)}
\int^{t_0}_t \exp{( {4\over \lambda}t')} A^2(t')\, dt' ,
\end{equation}
respectively, where $t_0$  is an arbitrary constant of
integration.
 
The  exact solutions (3.32) for $A(t)$ and (3.33) for
dynamically generated quark mass function $B(t_0,t)$
are not entire functions. The functions $A(t)$
and $B(t_0,t)$ have removable singularities at zero. In
addition, the dynamically generated quark mass function $B(t_0,t)$
also has algebraic branch points at $t=t_0$ and
at infinity. Apparently, these
unphysical singularities are due to ghost contributions.
The quark propagator may or may not be an entire function,
but in any
case the pole-type singularities should disappear. This is a
general feature of quark confinement and holds in any gauge. Our
solutions are of the
confinement-type since they have no pole singularities in the
whole complex $t$-plane on the real axis at any finite points.
Thus they satisfy the quark confinement criterion - disappearance
of the pole-type singularities in the corresponding quark Green's
functions.
 
   It is easy to see that $A(t)$ automatically has a correct
behaviour at infinity ( it approaches the free
quark propagator). In order to reproduce automatically the
correct behaviour of the dynamically generated quark mass function
at infinity, it is necessary to put $t_0=\infty$ in (3.33)
from the very beginning. Obviously in this case solution (3.33)
cannot be accepted at zero point ($t=0$), so that one needs
to keep the constant of integration $t_0$ arbitrary, but finite,
in order to obtain a regular, finite solution at zero point.
 
  The region $t_0 > t$ can be considered as nonperturbative,
whereas the
region $t_0 \le t$ can be considered as perturbative.
By approximating the full gluon propagator by its
deep IR asymptotics such as  $(q^2)^{-2}$ in the whole
range
$\left[ 0,  \infty \right) $,  we nevertheless obtain that our
solution  for the dynamical quark mass function $B(t_0, t))$
manifests
the existence of the boundary value momentum (dimensionless)
$t_0$ ,  which separates the IR (nonperturbative) region
from the intermediate and UV (perturbative) regions.
If QCD confines then a characteristic scale, at which
confinement and other nonperturbative
effects become essential, must exist. The
arbitrary constant of integration $t_0$  should be related to a scale
at which nonperturbative phenomena (such as confinement and
DCSB) dominate.
 
On one hand, playing a role of the UV cut-off ($t_0 =
\Lambda$), the
arbitrary constant of integration $t_0$ thereby prevents our quark
propagator from having an imaginary part. This is consistent
with the idea that a confined particle should have no imaginary
part [9]. On the other hand, this makes it possible to eliminate
the influence of the above mentioned unphysical singularities
which come from
the solutions to the quark SD equations (due to necessary ghost
contributions) on the $S$-matrix elements reproducing physical
quantities.
 
Thus, in order to obtain numerical values
of any physical quantity, e.g. the pion decay
constant (see below), the integration over the whole range
$\left[ 0, \  \infty \right] $
reduces to the integration over the nonperturbative region
$ \left[ 0, \  t_0 \right], $  which determines the range of
validity of the deep IR asymptotics (3.1) of the full gluon
propagator and consequently the range of validity of the
corresponding solutions (3.32) and (3.33) for the IR piece of
the full quark propagator.
 
  Here we ought to emphasize that the main contribution to
DCSB, as well as to the values of the pion decay constant and
other physical quantities, comes from the nonperturbative region
(large distances), whereas the contributions from the
short and intermediate distances (perturbative region)
can only be treated as perturbative corrections. We have
confirmed this physically reasonable assertion numerically
in the present paper. Despite having a formally
correct asymptotic behaviour at infinity, solutions (3.32)
and (3.33) for $B(t_0,t)$ (at $t_0=\infty$) cannot be responsible
for the UV asymptotic behaviour of the exact quark propagator. It
is better
to say that they do not spoil the UV asymptotics of the exact
quark propagator  rather than they reproduce
it. In order to reproduce correctly the UV asymptotics of the
exact
quark propagator, it is necessary to make renormalization group
improvements over the decomposition of the full gluon propagator
(3.6). In other words, one needs to take into consideration the
short distance contributions to the full gluon propagator.
As it was mentioned above, they are not important for the
numerical calculations, but from a conceptual point of view
they are certainly important. Unfortunately, the
way this can be done
self-consistently, i.e. taking into account gauge covariance,
ghost contributions, etc. is not yet known. We will discuss this
problem briefly at the end of this paper (Section 10).
 
  From Eqs.(3.32-3.33) it can easily be seen
that our quark propagator is nonperturbative and
"gauge-invariant", it is free of ghost complications (since it
does not depend explicitly on a gauge-fixing parameter and
the ghost self-energy at zero point) and it has no pole
 at any finite point on the real axis in
 the complex t-plane (confinement solution).
 Our quark propagator is also regular at zero point and
approaches the free propagator at infinity. So the
proposed quark propagator has only one $inevitable$
defect, it explicitly reproduces unphysical singularities due
to ghost contributions. A solution to this
problem, in our approach, is
to integrate the $S$-matrix elements of physical
quantities over the finite nonperturbative region ($t < t_0$)
whose size is determined by the confinement scale $\Lambda_c$
[16, 46] (see also next sections).
 
  Now come the main conclusions. Firstly, if the
enhancement of the zero modes of vacuum
fluctuations (3.7-3.9) takes place indeed then the quark Green's
function, reconstructed on the basis of this effect, has no poles.
In other words, the enhancement of the zero modes at the
expense of the virtual gluons alone removes nevertheless a
"single"
quark from the mass-shell (quark confinement theorem). Secondly,
a chiral symmetry violating part of the quark
propagator in this case is automatically generated. From the
obtained system (3.29-3.30) explicitly follows that it does
not allow a chiral symmetry preserving solution
($m_0 = B(t) = 0,  A(t) \ne 0 $). So a chiral symmetry violating
solution ($m_0 = 0, B(t) \ne 0,  A(t) \ne 0 $) for the quark
SD equation $is \ required$ and this is in
intrinsic connection with the absence of the pole
singularities in this solution. As was it emphasized above,
this solution (3.32-3.33) ( the  "ground
solution") will be used to compute basic chiral QCD
parameters, as well as the vacuum energy density, in the remaining
part of this paper.

\section{The vacuum energy density and gluon condensate }

    Any correct nonperturbative model of quark confinement and
DCSB necessarily becomes a model of the ground state
of QCD, i. e. the nonperturbative vacuum.
The effective potential method for composite operators [5, 6, 47]
allows one to investigate the vacuum of QCD since in the
absence of external sources the effective potential is nothing
but the vacuum energy density.
In this section we will evaluate the vacuum energy density
within the ZME effect (3.1) in QCD.
 
 The effective potential at one-loop is [47]
\begin{eqnarray}
V(S, D) &=& V(S) + V(D) =  \nonumber\\
&-& i \int {d^np \over {(2\pi)^n}}
Tr\{ \ln (S_0^{-1}S) - (S_0^{-1}S) + 1 \} \nonumber\\
&+& i { 1 \over 2} \int {d^np \over {(2\pi)^n}}
 Tr\{ \ln (D_0^{-1}D) - (D_0^{-1}D) + 1 \},
\end{eqnarray}
where $S(p)$ (2.2), $S_0(p)$ (3.23) and $D(p)$ (3.6), $D_0(p)$ are
the full, free quark and gluon propagators, respectively. The
trace over space-time
and colour group indices is implied to be taken but they
are suppressed for simplicity's sake  in
this equation. Let us recall now that the free gluon propagator
can be obtained from (3.6) by setting simply $d(-q^2, a) = 1$.
The effective potential is normalized as follows
\begin{equation}
V(S_0, D_0) = V(S_0) = V(D_0) = 0.
\end{equation}
 
In order to evaluate the effective potential,
let us use the well-known expression
\begin{equation}
 Tr \ln (S_0^{-1}S) = 3 \times \ln det (S_0^{-1}S) =
 3 \times 2 \ln p^2 \left[ p^2 A^2(-p^2) - B^2(-p^2) \right],
\end{equation}
where
\begin{equation}
p^2 A^2(-p^2) - B^2(-p^2)  = \sqrt det[-iS(p)].
\end{equation}
The factor 3 comes from the trace over quark colour
indices. Going over to Euclidean space
 ($d^4p \rightarrow i d^np, \quad p^2 \rightarrow -p^2$ )
and dimensionless variables (3.28), we finally obtain,
after some algebra, ($n=4$)
\begin{equation}
V(S) = {{6 \mu^4 \pi^2}\over {(2 \pi )^4}}
\int \limits_0^{t_0} dt\, t\, \{ \ln t\left[ t A^2(t) +
B^2(t_0, t)\right] + 2t A(t) + 2\},
\end{equation}
where we introduced the UV cutoff ($\Lambda$) which should be
identified with the arbitrary constant
of integration $t_0$, as it was discussed in the previous section
[48].
The explicit expressions for $A(t)$ and $B^2(t_0, t)$ are given
by (3.32) and (3.33) respectively.
 
In a rather similar way, for the gluon part one has
\begin{equation}
 Tr \ln (D_0^{-1}D) = 8 \times \ln det (D_0^{-1}D) =
 8 \times 4 \ln \left[ {3 \over 4 }d(-p^2) + {1 \over 4 } \right],
\end{equation}
where the factor 8 is due to the trace over the gluon colour
indices.
After doing some similar algebra and on account of (3.7), one
obtains in Euclidean space
\begin{equation}
V(D) = - {{16 \mu^4 \pi^2}\over {(2 \pi )^4}}
\int \limits_0^{t_0} dt\, t\, \{ \ln (3 {\lambda \over t}
+ 1 ) - {3 \lambda \over 4t} + b\},
\end{equation}
where $b={3 \over 4} - 2 \ln 2$. In derivation of (4.7)
we extracted the coupling constant $\lambda$ from $ d(-p^2)$ in
(3.7).
 
 Because of the normalization of the effective potential (4.2),
the vacuum energy density now should be defined as follows
\begin{equation}
\epsilon = V(S_0, D_0) - V(S, D) = - V(S,D).
\end{equation}
This means that the perturbative vacuum is normalized to zero
bacause of (4.2) and hence
\begin{equation}
\epsilon = \epsilon_q + \epsilon_g
\end{equation}
with
\begin{equation}
\epsilon_q = - V(S), \qquad \epsilon_g = - V(D),
\end{equation}
where $V(S)$ and $V(D)$ are given by (4.5) and (4.7),
respectively.
 
For the numerical calculation of the vacuum energy density (4.9),
along with other chiral QCD quantities below, it is convenient to
define some new variable and parameters by the relations
\begin{equation}
{2 \over \lambda } t = z, \qquad {2 \over \lambda } t_0 = z_0,
\qquad  t_0 = { k^2_0 \over \mu^2 },
\end{equation}
in terms of which we recast (4.5), on account of (4.10), as
\begin{equation}
\epsilon_q = - {3 \over 8 \pi^2} k^4_0 z^{-2}_0
\int \limits_0^{z_0} dz\, z\, \{ \ln z\left[ z g^2(z) +
B^2(z_0, z)\right] - 2z g(z) + 2\},
\end{equation}
where
\begin{equation}
g(z) =  z^{-2} [\exp{(-z)} -1 + z]
\end{equation}
and
\begin{equation}
B^2(z_0, z) =  3 \exp{(-2z)}
\int \limits_z^{z_0} {\exp{(2z')} g^2(z')\,dz'},
\end{equation}
so that
\begin{equation}
{\lambda \over 2} A(t) = - g(z), \quad {\lambda \over 2} B^2 (t_0,
t) = B^2 (z_0, z).
\end{equation}
This expression will be used for the numerical calculation
of the vacuum energy density contribution of the
confining quarks with dynamically generated masses.
It is instructive to write down the initial
system of equations (3.29-3.30) in terms of the new variables and
parameters defined in (4.11). In this case it  reads as
\begin{equation}
 g'(z)= - [{2 \over z} + 1 ] g(z)
+ { 1 \over z } + \tilde{m_0} B(z),
\end{equation}
\begin{equation}
B'(z) = -{3\over 2} g^2(z) B^{-1}(z) - [\tilde{m_0} g(z) + B(z)],
\end{equation}
where $\tilde{m_0} = m_0 ({2 \over \lambda})^{1/2}$. It is easy
to check that (4.13) and (4.14) are solutions to this system in
the chiral limit $\tilde{m_0} = 0$.
 
The vacuum energy density of the nonperturbative gluon
contributions (4.7), on account of (4.10), in the same variables,
is
\begin{equation}
\epsilon_g =  {1 \over \pi^2} k^4_0 z^{-2}_0 \times I_g(z_0)
\end{equation}
where
\begin{eqnarray}
I_g(z_0) &=&
\int \limits_0^{z_0} dz\, z\, \{ \ln (1+ {6 \over z})
 - {3 \over 2z } + b\}  \nonumber\\
&=& {1 \over 2} z^2_0 \ln (1
+ {6 \over z_0}) - 18 \ln (1 + { z_0 \over 6}) + {3 \over 2} z_0
+ {1 \over 2} b z^2_0.
\end{eqnarray}
Here one important remark is in place. The vacuum energy density,
$\epsilon_g$, does not vanish at $z_0 \rightarrow \infty$ as it
should because of (4.2). Thus it needs an additional
regularization in this limit. From (4.19) it follows that the
term containing the constant $b$ should be substracted from this
expression. So the regularized vacuum energy density should be
calculated with the relation (4.18) which now becomes
\begin{equation}
\epsilon_g = - {1 \over \pi^2} k^4_0 z^{-2}_0 \times
\left[ 18 \ln (1 + { z_0 \over 6}) - {1 \over 2} z^2_0
\ln (1 + {6 \over z_0}) - {3 \over 2} z_0 \right].
\end{equation}
The vacuum energy density (4.12)
automatically disappears at $z_0 \rightarrow \infty$, so it
does not requires an additional regularization, neither do the
other chiral QCD parameters (see below).

The vacuum energy density
is important in its own right as the main characteristics of the
nonperturbative vacuum of QCD. Futhermore it assists in
estimating such an important phenomenogical parameter as
the gluon condensate, introduced within the QCD sum rules approach
to resonance physics [49]. Indeed, because of the Lorentz
invariance,
\begin{equation}
\langle{0} | \Theta_{\mu\mu} | {0}\rangle = 4 \epsilon
\end{equation}
holds where $\Theta_{\mu\mu}$ is the trace of the energy-momentum
tensor and $\epsilon$ is the sum of all possible contributions to
the vacuum energy density, in particular is the sum of
$\epsilon_q$ and $\epsilon_g$ at least.
According to QCD this is equal to
\begin{equation}
\Theta_{\mu\mu} = {\beta(\alpha_s) \over 4 \alpha_s}
G^a_{\mu\nu} G^a_{\mu\nu}
\end{equation}
in the chiral limit ($G^a_{\mu\nu}$ being the gluon field
strength tensor) [50]. The CS-GML function $\beta(\alpha_s)$, up
to terms of order $\alpha^3_s$, is:
\begin{equation}
\beta(\alpha_s) = - { 9 \alpha^2_s \over 2 \pi} + O(\alpha^3_s).
\end{equation}
Sandwiching (4.22) between vacuum states and on account of
(4.23), one obtains
\begin{equation}
\langle{0} | \Theta_{\mu\mu} | {0}\rangle = - {9 \over 8}
\langle{0} | {\alpha_s \over \pi} G^a_{\mu\nu} G^a_{\mu\nu} | {0}\rangle
\end{equation}
where
$\langle{0} | {\alpha_s \over \pi} G^a_{\mu\nu} G^a_{\mu\nu} |
{0}\rangle$ is nothing but the gluon condensate [49].
From (4.21) it follows
\begin{equation}
\langle{0} | {\alpha_s \over \pi} G^a_{\mu\nu} G^a_{\mu\nu} |
{0}\rangle = - {32 \over 9} \epsilon
\end{equation}
 The weakness of this derivation is, of course, relation (4.23)
which holds only in the perturbation theory. Unfortunately, a
calculation of the CS-GML function $\beta(\alpha_s)$ beyond the
perturbation theory is not yet known. In any case, it would be
enlightening to estimate the gluon condensate with the help
of (4.25) and
on account of (4.9), which is the sum of the contribution from
the confining
quarks with dynamically generated masses (4.12) and the
nonperturbative gluons
contribution (4.20). This can also
be expressed as a function of the dynamically generated quark
mass (see (5.16) below). To this end, all is needed is to replace
the combination $k^4_0 z^{-2}_0$ in (4.12) and (4.20)
by the combination $m^4_d B^4 (z_0, 0)$.

\section{ The basic chiral QCD parameters}

Because of its especially
small mass, the pion is the most striking example of the Goldstone
realization of the chiral symmetry $SU(2)_L \times SU(2)_R$.
The pion decay constant  $F_\pi$ is an
important constant in Nature.
On one hand it determines the rate of the pion
semileptonic decays $\pi^{\pm} \rightarrow e^{\pm} \nu_e,
\mu^{\pm} \nu_\mu$,
on the other hand, beside the quark condensate and the dynamically
generated quark mass, it is one of the three most important chiral
QCD parameters that determine the scale of chiral dynamics. All
other parameters in chiral QCD can be expressed in terms of
these independent basic parameters.

\subsection{The pion decay constant in the current algebra
            representation}

  In order to show explicitly how the CHPTq works, and for further
aims, let us apply the CHPTq directly to the axial-vector matrix
element which determines such an important physical parameter
as the pion decay constant. As it is well known, the pion
decay constant $F_{\pi}$ is defined in the current algebra (CA) as
\begin{equation}
\langle{0}| J^i_{5\mu}(0) |{\pi^j(q)} \rangle = iF_{\pi}q_\mu
\delta^{ij} .
\end{equation}
(Here the normalization $F_{\pi} = 92.4 MeV$ is used [51]).
Clearly, this matrix element can be written in terms of the pion-
quark-antiquark proper vertex and quark propagators as
\begin{equation}
 iF_{\pi}q_{\mu} \delta^{ij} = \int {d^4p\over {(2\pi)^4}} Tr\bigl\{
 \left({\lambda^i \over 2}\right)\gamma_5\gamma_{\mu}
S(p+q)G^j_5(p+q,p)S(p)\bigr\}.
\end{equation}
The trace is understood over the Dirac and colour
indices. To get an expression for $F_{\pi}$  one has
to differentiate Eq.(5.2) with respect to the external momentum
$q_{\nu}$  and then set $q=0$ (Fig. 2). Let us
denote the first term in Fig. 2 by  $D^{i j}_{\mu \nu}(a)$
and the second one by $D^{i j}_{\mu \nu}(b)$,
respectively. Thus, in general, one has
\begin{equation}
iF_{\pi} = D(a) + D(b) = \delta^{ij} g_{\mu\nu}
[D^{ij}_{\mu\nu}(a) + D^{ij}_{\mu\nu}(b)] ,
\end{equation}
   By taking into account the BS pion wave function, up
to terms of order $q$ given by (2.18-2.19) and (2.2) with the
substitution $p \rightarrow p+q$ and also using the
Taylor expansions (2.15-2.16), this expression
can be easily evaluated. Now it is
clear that in order to investigate analytically, and calculate
the pion decay constant ( and many other important physical
parameters) numerically, our parametrization of the
full quark propagator (2.2) is technically much more
convenient than (2.3) is. This parametrization is also
preferable for investigations of the confinement
properties of the quark propagator [12, 13, 16].
  Going over to the Euclidean space
$(d^4 p \rightarrow i d^4 p,\ p^2 \rightarrow - p^2)$
and using dimensional variables (3.28), one finally obtains
that the sum of two diagrams (5.3) should read as follows
\begin{eqnarray}
F^2_{CA} = {{12\pi^2}\over{(2\pi)^4}}\mu^2
   {\int}^{\infty}_0 dt\, t \{ -\overline B(t)
   [AB+{1\over 2}t(A'B-AB')] \nonumber\\
   - {3\over 4}tABR_{11}(t) + {1\over 4}R_6(t)(E-3B^2) \},
\end{eqnarray}
 
  The first line of this equation describes the contribution
coming from the first diagram in Fig. 2, while the second line
is the contribution from the second diagram.
The contribution from the second diagram completely
disappears if one uses the pion BS bound-state
amplitude, restored from the axial-vector WT identity in a
self-consistent way, as a residue at pole (2.22-2.26).
It is the contribution of the first diagram in
Fig. 2 that leads to the well-known expression of
Pagels-Stokar-Cornwall (PSC) for the pion decay constant in
the chiral QCD [5, 8, 23, 24] (see our paper [20]).
Here and below the primes denote differentiation with
respect to the dimensionless Euclidean momentum variable $t$,
$A=A(t), B=B(t)$ and quantities with overline are shown in (3.27)
on account of (3.28). We denote the pion
decay constant $F_{\pi}$, in the chiral limit of the CA
representation,
by $F_{CA}$. This is a general CHPTq
expression for the pion decay constant in the CA representation,
obtained from the BS bound-state amplitude, which is restored
from the corresponding axial-vector WT identity in
a self-consistent way, up to terms of order $q$.
 
   The main problem now is to find a good
nonperturbative ansatz (mentioned above) for the arbitrary
form factors $R_j(t) (j=6,11)$ in the IR region.
In nonperturbative calculations these terms cannot be ignored
by saying formally they are of order $g^2$,  as it was done in the
perturbative treatments [5, 8, 23, 24]. In connection with this,
let us point out that the difference between the vector and
axial-vector currents disappears in the chiral limit.
For this reason let us assume [46]
that the IR finite quark-gluon vector vertex function at
zero momentum transfer (3.25-3.26) is
a good approximation to the regular piece of the
axial-vector vertex at zero momentum transfer in the chiral limit
(2.20-2.21), which in Euclidean space becomes
\begin{eqnarray}
G_1(p^2) &=& - \overline A (p^2) - R_6(p^2)  \nonumber\\
G_2(p^2) &=&  2 \overline B'(p^2) \nonumber\\
G_3(p^2) &=&  2 \overline A'(p^2) \nonumber\\
G_4(p^2) &=& - R_{11}(p^2).
\end{eqnarray}
A fortunate feature that admits to exploite a partial analogy
between the
vector and axial-vector currents in the chiral limit for the
flavor non-singlet channel is that the contribution
to the pion decay constant in the CA representation (5.4),
coming from the second diagram in Fig. 2, does not depend on the
form factor $G_2(p^2)$ at all.  In this case the analogy between
(3.26) and (5.5) becomes complete and one obtains
\begin{equation}
R_6(p^2) = 0, \qquad
R_{11}(p^2) = - {{A^2(p^2)B^{-1}(p^2)} \over {E(p^2)}}.
\end{equation}
Certainly we cannot prove  these relations, though it will be
shown later (Section 6) that this dynamical assumption
(nonperturbative ansatz) numerically works very well and it
leads to a self-consistent calculation scheme
for all chiral QCD parameters thereby, justifying it once again.
 
  Using the quark SD equation of motions (3.29-3.30) for the
ground solution, taking (5.6) into account, after
some algebra one finally arrives at the following result for the
pion decay constant in the CA representation
\begin{equation}
F^2_{CA} = {{12\pi^2}\over{(2\pi)^4}}\mu^2 {1 \over \lambda }
     {\int}^{t_0}_0 dt\, {{tB^2(t_0,t)}\over
      {\{tA^2(t) + B^2(t_0,t)\}}} ,
\end{equation}
where we introduced a cut-off which should be identified with the
arbitrary constant of integration $t_0$, as it was discussed in
Section 3.
 
In terms of new parameters and variable (4.11), we recast
(5.7) as follows
\begin{equation}
F^2_{CA} = {3\over {8\pi^2}} k_0^2 z_0^{-1}
             \int^{z_0}_0 dz \,{ zB^2(z_0,z) \over
             {\{zg^2(z) + B^2(z_0,z)\}}}.
\end{equation}
The expression (5.8) will be used for the numerical
calculation of the pion decay constant in the CA representation.
This can also be expressed
as a function of the dynamically generated quark mass $m_d$
(see the next section) and the dimensionless parameter $z_0$, i.e.
\begin{equation}
F^2_{CA} = {3\over {8\pi^2}} m_d^2 B^2(z_0,0)
             \int^{z_0}_0 dz \,{ zB^2(z_0,z) \over
             {\{zg^2(z) + B^2(z_0,z)\}}} ,
\end{equation}
where $B^2(z_0,0)$ is given by (4.14) at $z=0$. This
expression clearly shows that a nonzero pion decay constant in the
chiral limit is possible only if dynamical quarks
exist depending, in turn, on the formation of quark
condensates (see below).
 
 Let us note that in QCD there exists an
exact representation of the pion
decay constant  $F_{\pi}$,  derived by Jackiw and Johnson (JJ)
in Ref. 4 (see also Refs. 8, 20, 23, 24). In comparison with
the CA representation their representation is free
from overlapping divergences but it requires an explicit
investigation of the corresponding BS scattering kernel. A
numerical investigation of the chiral QCD in the JJ
representation will be performed elsewhere.

\subsection{The dynamically generated quark mass}

  It is worthwhile to discuss the relation between the effective
quark mass $m_{eff}$ and the dynamically generated quark mass
$m_d$ in our model.
Let us define $m_{eff}$ as the inverse of the full quark
propagator (2.3) at the zero point [9, 16, 20]
\begin{equation}
m_{eff} = [iS(0)]^{-1} = \overline B(0).
\end{equation}
Let us define the dynamically generated quark mass $m_d$
in a similar way, namely
\begin{equation}
m_d = [iS_{ch}(0)]^{-1} = \overline B_0(0),
\end{equation}
where $S_{ch}(0)$ denotes the full quark propagator in the chiral
limit $m_0 =0$. Evidently, here $\overline B(0)$ is the
constant of integration of the full quark SD equation and
$\overline B_0(0)$ is the same constant in the chiral limit.
 
 From the quark SD equation
\begin{equation}
[iS(p)]^{-1} = [iS_0(p)]^{-1} + \Sigma(p),
\end{equation}
it follows that
\begin{equation}
m_{eff} = m_0 + m_d .
\end{equation}
In the chiral limit ($m_0=0$), $m_{eff} = m_d$ and thus $m_d$
arises only because of the interaction between quarks and gluons
(quark mass generation). So, by our definitions, an effective
chiral quark means simply a dynamical quark.
 
  These definitions allow to exploit
an analogy with the free quark propagator at zero point. They
are also consistent with the renormalization
group analysis in the IR region [1, 16]. These definitions,
on one hand, generalize the constituent quark model propagator
\begin{equation}
S(p) = {i \over \hat p - m_q},
\end{equation}
where $m_q$ denotes constituent quark mass. On the other hand,
they are consistent with the solutions to the quark SD
equation which have no
poles (confinement-type solutions), reconstructed on the basis of
the ZME effect due to the nonperturbative
IR divergences, leading to the both quark
confinement and DCSB phenomena in QCD (see preceding section, our
paper [16] and references therein). In connection with (5.14)
it makes sense to note in advance that from our numerical results
(Section 6), it follows that $m_q$ differs little from $m_d$,
i.e. $m_q = m_d + \Delta$ in the chiral limit. We are able also to
propose an
explanation of the dynamical origion of this $\Delta$ on account
of the ZME effect in bound-state problems (see Section 8).
 
 Here a few remarks are in order. In gauge theories to extract
gauge-invariant physical information is not a straightforward
procedure. For example, in QED the electron Green's function is an
explicitly gauge-dependent quantity but the position of the pole
is gauge-invariant. In QCD this fails because of the quark
confinement phenomenon which, as it was mentioned above, is
understood as the disappearence of the pole singularities in the
quark
Green's function. Obviously definitions (5.10) and (5.11) assume
regularity also at the zero point. Though the effective
quark mass $m_{eff}$ is not an experimentally observable
quantity, it is desirable to find
such kind of solutions to the quark SD equations
in which the explicite dependence on a gauge-fixing parameter
disappears. In this sense $m_{eff}$ and $m_d$, defined by
(5.10) and (5.11) respectively, become gauge-invariant. As it was
discussed in Section 3, such a
nonperturbative quark propagator has been found within
our approach to QCD at large distances [12, 13, 16].
 
  As we mentioned in Section 2, the $\gamma_5$
invariance of the quark propagator is broken, as shown by relation
(2.5), and the measure of this breakdown is the double of the
dynamically generated quark mass function $2 \overline B(-p^2)$.
This quantity at zero, $2 \overline B(0)$, can be defined
as a scale of DCSB at the fundamental quark level [20].
In accordance with (5.11), let us denote it as
\begin{equation}
 \Lambda_{CSBq} = 2 \overline B(0) = 2 m_d ,
\end{equation}
while recalling that in the chiral limit $m_{eff}=m_d$. The
definitions
$m_{eff}$, $m_d$ and $\Lambda_{CSBq}$ have now direct physical
sense within the above mentioned solutions to the quark SD equation.
 
  Let us write down also the final result for the
dynamically generated nonperturbative quark mass (5.11), expressed
in terms of new parameters and variable (4.11)
within the ground solution (3.33)
\begin{equation}
m_d = k_0\bigl\{z_0 B^2(z_0,0)\bigr\}^{-1/2},
\end{equation}
where $B^2(z_0,0)$ is given by (4.14) at the zero point.
It should be clear now that the dynamically
generated quark mass is a constant of integration of the
corresponding SD equation and is, at the same time, one
of the
two independent parameters of our treatment (see Section 6).
  Let us also express the dimensionful
parameter $k_0$ as function of $m_d$ and $z_0$:
\begin{equation}
k_0 = m_d \bigl\{z_0 B^2(z_0,0)\bigr\}^{1/2},
\end{equation}
emphasizing that precisely this parameter determines a scale
at which nonperturbative effects take the leading role.

\subsection{The quark condensate}

  As it is well known, the order parameter of DCSB - quark
condensate
can also be expressed in terms of the quark propagator scalar
function $B(-p^2)$ (2.2). The definition is
\begin{equation}
{\langle \overline qq \rangle} =
{\langle 0| \overline qq | 0 \rangle}
=  \int {d^4p\over {(2\pi)^4}} Tr S(p).
\end{equation}
Let us write down the final result expressed
in the new variables (4.11) as follows
\begin{equation}
{\langle \overline qq \rangle}_0 = -
{3\over {4\pi^2}}k_0^3z_0^{-3/2}{\int}^{z_0}_0 dz\,{zB(z_0,z)},
\end{equation}
and as a function of $m_d$ as
\begin{equation}
{\langle \overline qq \rangle}_0 = -
{3\over {4\pi^2}} m_d^3 B^3(z_0,0) {\int}^{z_0}_0
dz\,{zB(z_0,z)},
\end{equation}
where for light quarks in the chiral limit
\begin{equation}
{\langle \overline qq \rangle}_0 \equiv
{\langle 0| \overline uu | 0 \rangle}_0 =
{\langle 0| \overline dd | 0 \rangle}_0 =
{\langle 0| \overline ss | 0 \rangle}_0
\end{equation}
by definition. Eq. (5.20) clearly shows that in our model
the existence of the dynamically generated quark mass, which is a
consequence of the enhancement of the zero modes, should be
certainly accompanied by the
formation of quark condensates and vice versa. Quark condensates
lead to the creation of the effective (dynamical) quark mass.
 
 There are also two chiral QCD parameters closely related to the
quark condensate.
  Let us consider the coupling constant of the pseudoscalar
density to the pion which is defined as
\begin{equation}
{\langle 0| \bar q i \gamma_5{\tau^i \over 2} q |
{\pi^j(0)} \rangle } = G_\pi \delta^{ij}.
\end{equation}
 
Similarly to the pion decay constant (5.2), the analytical
expression for this matrix element is
\begin{equation}
G_{\pi} \delta^{ij} =  \int {d^4p\over {(2\pi)^4}}
Tr \{ i \gamma_5 \left( {\tau^i \over 2} \right)
S(p)G^j_5(p,p)S(p) \},
\end{equation}
where the trace is understood over Dirac and color indices.
Taking into consideration (2.2) for the quark propagators,
(2.18-2.19) for the pion wave function $G_5^j(p,p)$ at zero
momentum transfer, one can easily evaluate this expression by
substitution
 $\lambda^j \rightarrow \tau^j$, as
\begin{equation}
G_\pi\times F_\pi =  12 i  \int {d^4p\over {(2\pi)^4}}
B(- p^2) = - \int {d^4p\over {(2\pi)^4}} Tr S(p).
\end{equation}
Comparing this with the definition of the quark condensate
(5.18), one obtains the well-known current algebra result
\begin{equation}
G_\pi\times F_\pi = - {\langle \overline qq \rangle}
\end{equation}
and in the chiral limit, we have
\begin{equation}
$$ G\times F = - 2 {\langle \overline qq \rangle}_0,
\end{equation}
where we denote $G_\pi$ in the chiral limit by G, similarly like
it was done
above with F. This formula will be used
for the numerical calculation of this constant.
 
  In the chiral perturbation theory at the hadronic level
(CHPTh, or, equvivalently, the effective field theory) [52, 53]
there is a second low energy constant $B$, determined by
\begin{equation}
{\langle \overline qq \rangle}_0 = - F^2 B.
\end{equation}
This measures the vacuum expectation value of the scalar densities
in the chiral  limit.
It is just this constant that
determines the meson mass expansion in the general case. Indeed,
in leading order (in powers of quark masses
and $e^2$) from the CHPTh, one obtains [52, 53]
\begin{eqnarray}
 M^2_{\pi^+} & = & (m^0_u + m^0_d) B  \\
 M^2_{K^+} & = & (m^0_u + m^0_s) B  \\
 M^2_{K^0} & = & (m^0_d + m^0_s) B .
\end{eqnarray}
Calculating independently the constant $B$ from (5.27), one then
will be able to estimate the current quark masses
$m^0_u, m^0_d$ and $m^0_s$ by using the experimental values
of meson masses [54] in (5.28-5.30).

\subsection{The Goldberger-Treiman relations at the quark and
              hadronic levels}

  The famous Goldberger-Treiman (GT) relation
(identity) at the fundamental quark level reads as follows
\begin{equation}
F_{\pi} \overline G(p,p) = 2 \overline B(-p^2),
\end{equation}
where $\overline G(p,p)$ denotes
the left-hand-side (lhs) of (2.18) in the limit $q=0$.
We omitted the minus sign which appears because of the
presence of $\gamma_5 \gamma_{\mu}$
instead of the standard $\gamma_{\mu} \gamma_5$.
This is an unrenormalized (for a unit axial-vector coupling
constant of the constituent quark, i. e. $g_A = 1$) GT
relation at the quark level
[7-9, 24] and it is exact in the chiral limit.
 
 On account of (5.11) at zero momentum, one obtains, by simply
identifying
$\overline  G(0,0)$ with $2 g^o_{{\pi}qq}$, as well as by
denoting here, for the sake of convenience, the pion decay
constant
in the chiral limit as $F^o_{\pi}$, instead of previous $F$, the
GT relation  (5.31) at the quark level as
\begin{equation}
F^o_{\pi} g^o_{{\pi}qq} = m_d ,
\end{equation}
where $g^o_{{\pi}qq}$ denotes
pion-quark  coupling  constant  in  the  chiral  limit.  This is a
standard  form  of  the  unrenormalized ($g_A=1$) GT  relation
at the quark level.
In order to recover the renormalized GT  identity,
it is necessary only to multiply the rhs of relation (5.32)
by  $g^o_A$, which is the  above mentioned corresponding
axial-vector coupling constant of the constituent quark in the
chiral limit, and replace $m_d$ by $m_q$ i.e.,
\begin{equation}
F^o_{\pi} g^o_{{\pi}qq} = m_q g^o_A,
\end{equation}
where we retained the same notations for the
renormalized quantities. This is a renormalized GT relation at the
fundamental quark level and it becomes exact in the chiral limit
[55]. For the numerical calculation of $g^o_{{\pi}qq}$
this relation (5.33) will be used, even though it
requires knowledge of $g^o_A$ which remains still
unknown in our calculation scheme at this stage. However, let us
put this problem aside until the end of this subsection.
 
   The standard GT identity [56, 57] for pion-nucleon physics
reads as
\begin{equation}
F_{\pi} g_{{\pi}NN} = M_N G_A.
\end{equation}
Experimental values of
the pion-nucleon coupling constant $g_{{\pi}NN}$ and the nucleon
axial-vector coupling $G_A$, as well as of the pion decay
constant and nucleon mass, are well known [51, 54, 57], namely
\begin{eqnarray}
g_{{\pi}NN} &=& 13.04 \pm 0.06, \qquad G_A = 1.257 \pm 0.003 \\
F_{\pi} &=& 92.4189 \ MeV,\qquad M_N = {1 \over 2} (M_p + M_n) =
938.92 \ MeV
\end{eqnarray}
Note that just these values, denoted as
$g_{{\pi}pn}$ and $\sqrt 2 f_{\pi} = 2 F_{\pi}$,
are quoted in Ref.57. From these experimental values it
follows that (in units of MeV)
\begin{equation}
F_{\pi} g_{{\pi}NN} = 1205.14 > M_N G_A = 1180.22
\end{equation}
Following the
paper of Marciano and Sirlin [57], let us consider the discrepancy
between the rhs and lhs of the GT relation (5.34) as
\begin{equation}
\Delta_{\pi} = -1 + {F_{\pi} g_{{\pi}NN} \over M_N G_A}= 0.02111,
\end{equation}
where we used numerical values from (5.35-5.36). So, in the
non-chiral case, the GT relation is not exact and it is expected
to hold up to (1-2)\% accuracy [57].
 
  In the chiral limit the GT relation becomes exact. This
means that the quantities entering the GT relation (5.34)
slightly change their numerical values in the chiral limit.
Denoting, in addition, the pion-nucleon coupling constant in
the chiral limit as $g^o_{{\pi}NN}$, as well as nucleon mass as
$M^o_N$, one obtains
\begin{equation}
F^o_{\pi} g^o_{{\pi}NN} = M^o_N G^o_A.
\end{equation}
This is the exact GT relation at the hadronic level (for
pion-nucleon physics) in the chiral limit. The above mentioned
discrepancy (5.38) should vanish in this
case. In agreement with the statement that the numerical values
of the physical quantities (those considered in this paper) in the
chiral limit should not exeed
their experimental values, and on account of (5.37), it follows
that the deviation of the lhs in the GT relation (5.34) is larger
than the deviation of the rhs from their experimental values.
 
  Unfortunately, at this stage, we are not able to independently
calculate the pure
nucleonic degrees of freedom in the chiral limit, in particular
$M^o_N$ and $G^o_A$. Though we might have taken these
values from their lattice calculation reported in Refs. 58
but we are not going to proceed this way because that would
undesirably complicate our calculation scheme and would not give
new information either. Note that some
serious problems with the chiral limit itself ("hard" chiral
logarithms), recently discovered in quenched lattice
calculations [59], prompted us not to use these results. For the
sake of transparancy of our calculation scheme, it completely
suffices to use the chiral value of the pion decay constant
as obtained by the CHPTh in Ref. 60, namely
\begin{equation}
F^o_{\pi} = (88.3 \pm 1.1) \ MeV.
\end{equation}
This value is chosen as unique input data
in our numerical investigation of chiral QCD.
The pion decay constant is a good experimental number since it is
a directly measurable quantity in contrast, for example,
to the quark condensate or the above mentioned pion-quark coupling
constant. For this reason our choise (5.40) as input data opens up
the possibility of realibly estimating the deviation of the
chiral
values from their "experimental" phenomenologically determined
values of various physical quantities which can not be directly
measured. Thus to assign definite values to the physical
quantities in the chiral limit is a rather delicate question. At
the same time it is a very important theoretical
limit which determines the dynamical structure of low-energy QCD.
Note that from now on we will use
only the central value of the pion decay constant in the chiral
limit as given by (5.40).
 
  Let us now return to the GT relation (5.33). In
QCM [32, 61] the following relation holds true on account of (5.35).
\begin{equation}
g_A = {3 \over 5} \times G_A = 0.7542.
\end{equation}
This value is usually considered as the
phenomenological ("experimental") value of this quantity (see,
however, the second paper in Refs. 55).
Let us assume that relation (5.41) is valid in the
chiral limit too. Then from (5.33) and (5.39) one obtains
\begin{equation}
g^o_{{\pi}qq} = {3 \over 5} {m_q \over M^o_N} g^o_{{\pi}NN}.
\end{equation}
Having approximated the chiral
values of $M^o_N$ and $g^o_{{\pi}NN}$ with their experimental
values (5.36) and (5.35) respectively, this expression yields the
value for the pion-quark coupling constant.

\section{Numerical results}

  Let us begin this section with the discussion of one of the most
interesting features of DCSB. As it was explicitly shown above,
there
are only five independent quantities by means of which all other
chiral QCD parameters can be calculated. For the sake of
convenience, let us write down them together.
\begin{equation}
F^2_{CA} = {3\over {8 \pi^2}}k_0^2z_0^{-1}
             \int^{z_0}_0 dz \,{ zB^2(z_0,z) \over
             {\{zg^2(z) + B^2(z_0,z)\}}} ,
\end{equation}
\begin{equation}
m_d = k_0\bigl\{z_0 B^2(z_0,0)\bigr\}^{-1/2},
\end{equation}
\begin{equation}
{\langle \overline qq \rangle}_0 = -
{3\over {4\pi^2}}k_0^3z_0^{-3/2}{\int}^{z_0}_0 dz\,{zB(z_0,z)},
\end{equation}
\begin{equation}
\epsilon_q = - {3 \over {8 \pi^2}} k^4_0 z^{-2}_0
\int \limits_0^{z_0} dz\, z\, \{ \ln z\left[ z g^2(z) +
B^2(z_0, z)\right] - 2z g(z) + 2\},
\end{equation}
\begin{equation}
\epsilon_g = - {1 \over \pi^2} k^4_0 z^{-2}_0 \times
\left[ 18 \ln (1 + { z_0 \over 6})
       - {1 \over 2} z^2_0 \ln (1 + {6 \over z_0})
       - {3 \over 2} z_0 \right].
\end{equation}
recall that $B^2(z_0,z)$ and $g(z)$ are given by (4.14)
and (4.13), respectively. The
definition (5.15) for the DCSB scale is
\begin{equation}
 \Lambda_{CSBq} = 2 m_d.
\end{equation}
These final expressions are going to
be used to calculate the chiral QCD parameters. They
depend only on two independent quantities, namely the mass scale
parameter $k_0$ and the constant
of integration of dynamical quark SD equation of motion $z_0$.
However, from (6.2) we get that information on the
parameter
$z_0$ must be extracted  from $m_d$ and the initial mass
scale parameter $k_0$,
which characterizes the region where confinement, DCSB and
other nonperturbative effects are dominant.
The second indepent parameter of the ground
solution $z_0$ is given in terms of $k_0$ and $m_d$.
Despite the fact that in our consideration the initial mass scale
parameter $\mu$ (characterizing the scale of nonperturbative
effects) has been introduced "by hand", such a transformation of
pair of independent parameters $k_0$ and $z_0$ into the pair of
$k_0$ and $m_d$ is
also a direct manifestation of the phenomenon of the "dimensional
transmutation" [62]. This phenomenon occurs whenever a massless
theory acquires masses dynamically. It is a general feature of
spontaneous symmetry breaking in field theories.
 
Our calculation scheme is self-consistent because we calculate
$n=5$ independent physical quantities by means of $m=2$ free
parameters, which possess clear physical sense. So condition of
self-consistensy $n > m$ is satisfied. The general behaviour of
all our parameters, given by relations (6.1-6.5), are
shown in Figs. 3-7.
In our model one may calculate all the chiral QCD
parameters (not only those considered here but others as well!) at
any requested combination of $m_d$ and $k_0$. However, in order to
analyse our numerical results, it is necessary to set a
scale at which it should be done. We set a scale by two, at
first sight different, ways but these lead to
almost the same numerical results.

\subsection{Analysis of the numerical data at a
             scale of DCSB at the quark level}

  There is a natural
scale in our approach to DCSB. At the
fundamental quark level the chiral symmetry is spontaneously broken
at a scale $\Lambda_{CSBq}$ defined by (6.6). We may then
analyse our numerical data at a scale at which DCSB at the
fundamental quark level occurs. For this aim, what is needed is
only to simply identify mass scale parameter $k_0$
with this scale $\Lambda_{CSBq}$, i.e. to put
\begin{equation}
k_0 \equiv \Lambda_{CSBq} = 2 m_d.
\end{equation}
Now one can uniquely determine the constant
of integration of the quark SD equation. Indeed, from
(6.2) and on account of (6.7), then it immediately follows that
this constant is equal to
\begin{equation}
z_0 = 1.34805.
\end{equation}
  From the pion decay constant
in the chiral limit (5.40), chosen as input data, and on account
of (6.8) and (6.7), from (6.1) one obtains a numerical value for
$k_0$ (see Table 1). This means that all physical
parameters considered in our paper are uniquely determined.
The results of our calculations are displayed in Table 1.
 
  It is plausible that our approach makes it
possible to intrinsically compare numerical results of different
approaches with each other. For example, of the CHPTh [52, 53, 60]
with those of QCD sum rules [49, 63] and the CQM
[32, 61, 64] as well as lattice approach [58] and vice versa.
In the most simplest way this can be done
by setting a scale based on definition (6.7)
(calculation scheme A). To achieve this goal one should
chose input data from the corresponding approach and
then proceed as it was described above. We do not present
these calculations. It will be
instructive to display our numerical results when the
chiral value of the pion decay constant is approximated by
its experimental value, as advocated in Refs. 51 and 57, namely
\begin{equation}
F^o_{\pi} = 92.42 \ MeV,
\end{equation}
as well as by the standard value
\begin{equation}
F^o_{\pi} = 93.3 \ MeV.
\end{equation}
In accordance with the above mentioned, this allows for
estimating the deviation of the chiral
values of the physical parameters, which can not be directly
mesuared, from their phenomenologically estimated ("experimental")
values. For the above calculated parameters these results are
also shown in Table 1.
 
The estimate of quark condensate in Refs. 49 and 63,
\begin{equation}
{\langle \overline qq \rangle}_0^{1/3}  = -(225 \pm  25) \ MeV
\end{equation}
is in good agreement with our values.
QCD sum rules give usually the numerical values of physical
quantities, in particular the quark condensate, approximately,
within an accuracy of (10-20)\% (see, for example Ref. 65).
 
   Our values for the current quark masses are also in good
agreement with recent estimates from hadron mass splittings [66]
\begin{eqnarray}
m^0_u = (5.1 \pm 0.9)  \ MeV,   \nonumber\\
m^0_d = (9.0 \pm 1.6)  \ MeV,   \nonumber\\
m^0_s = (161 \pm 28) \ MeV
\end{eqnarray}
and QCD sum rules [67]
\begin{eqnarray}
m^0_u = (5.6 \pm 1.1)  \ MeV,   \nonumber\\
m^0_d = (9.9 \pm 1.1)  \ MeV,   \nonumber\\
m^0_s = (199 \pm 33) \ MeV,
\end{eqnarray}
see also reviews [68]. Note that the agreement of
our values (Table 1) with the QCD sum rules values (6.13) is
slightly better than with those of (6.12) obtained from hadron
mass splittings.
 
 In order to obtain the numerical values of the
pion-quark coupling constant (5.42), we approximated
$m_q$ in (5.42) by the numerical values of $m_d$ (Table 1).
Here it is worth noting in
advance that from our model of quark confinement (see discussion
in Section 8) it follows that $m_q$ differs
little from $m_d$; so, making no big mistake, one can
simply use $m_d$ instead of $m_q$ in (5.42). Moreover, doing so
one arrives at the conclusion that the chiral perturbation theory
with (5.40) and the CQM with the value for the constituent
quark mass $m_q=362 \ MeV$ advocated by Quigg [64] are nearly in
one-to-one correspondence n our calculation scheme A (see
Table 1).
 
  The phenomenological analysis of QCD sum rules [49] for the
numerical value of the gluon condensate implies
\begin{equation}
\langle{0}|{\alpha_s \over
\pi}G^a_{\mu\nu}G^a_{\mu\nu}|{0}\rangle \simeq  0.012 \ GeV^4,
\end{equation}
and using then again (4.25), one obtains the vacuum energy density
\begin{equation}
\epsilon \simeq -0.003375 \ GeV^4
\end{equation}
with QCD sum rules [49]. In the random instanton
liquid model (RILM) [69] of the QCD vacuum, for a dilute
ensemble, one has
\begin{equation}
\epsilon =-{ 9 \over 4} \times 1.0 \ fm^{-4} \simeq -0.003411 \ GeV^4.
\end{equation}
The estimate, with QCD sum rules,
of the gluon condensate can be changed within a factor of two [49].
We trust our numerical results for the vacuum energy density
much more than those of the gluon condensate.
The former estimate was obtained on the basis of the completely
nonperturbative ZME model of the vacuum of QCD while the latter
one was obtained on account of the perturbative solution for the
CS-GML $\beta$-function (4.23). In order to realibly calculate
gluon condensate from (4.22), it is necessary to calculate the
CS-GML $\beta$-function in the nonperturbative ZME model of quark
confinement and DCSB. This calculation is not straightforward but
we undertake the task. Let us also emphasize the important
fact that our calculation of the vacuum energy density is a
calculation from first principles while in the RILM the parameters
characterizing the vacuum, the instanton size $\rho_0 = 1/3 \ fm$
and the "average separation" $R= 1.0 \ fm$ were chosen
to precisely reproduce traditional (phenomenologically estimated
from QCD sum rules) values of quark and gluon condensates,
respectively.
 
We reproduce values (6.14-6.16), which are due to
the instanton-type fluctuations only, especially well
when the pion decay constant in the chiral limit was
approximated by its experimental value. Moreover, our numerical
results clearly show that the contribution to the vacuum energy
density by the confining quarks (with dynamically generated
masses), $\epsilon_q$, is approximately equal to $\epsilon_g$,
which comes from the nonperturbative gluons. It is
well known that in the chiral limit (massless
quarks) tunneling is totally suppressed, i.e. the contribution
to the vacuum energy density of the instanton-type fluctuations
vanishes. It will be restored again
in the presence of DCSB [70, 71]. Thus, in
general, the total vacuum energy density should be the sum of
these three quantities (see discussion in Section 8).
 
To set a scale by the way described in this subsection has the
advantage that it
is based on the exact definition (6.7) for a scale at which
analysis of the numerical data must be done. This is a scale
responsible for DCSB at the fundamental quark level. In
general it is not obvious at all
that this scale $\Lambda_{CSBq}$ and the scale at which
quark confinement occurs, $\Lambda_c$, should be of the same
order of magnitude. Moreover the information about $\Lambda_c$,
at which quark confinement occurs at the fundamental microscopic
level, is hidden in this scheme of calculation.
In order to clear up the origin of $\Lambda_c$ and its relation to
$\Lambda_{CSBq}$, let us set a scale in the way described in the
next subsection.

\subsection{Analysis of the numerical data at the
            confinement scale}

  As noted above, there exists only one scale in our model,
denoted first by $\mu$ and then by $k_0$ (to
distinguish between
the nonperturbative phase and the perturbative one), which is
responsible for all nonperturbative effects in QCD at large
distances. If there is a close connection between quark
confinement and DCSB (and we believe that this is so) then the
scale of DCSB at the fundamental quark level $\Lambda_{CSBq}$
(6.6) and the confinement scale
$\Lambda_c$ should be, at least, of the same order of magnitude.
This is in agreement with Monte Carlo
simulations on the lattice which show that the deconfinement phase
transition and the chiral symmetry restoring phase transition
occur approximately at the same critical temperature [72, 73],
confirming thereby the close intrinsic link between these
nonperturbative phenomena.
 
   Unfortunately, the exact value of neither $m_d$ or of
$k_0$ is known. For this reason, let us first
reasonably assume that for the dynamically generated
quark masses
\begin{equation}
 300  \le   m_d \le 400  \ (MeV)
\end{equation}
is imposed but they otherwise remain arbitrary.
We believe that this interval covers all possible realistic
values used for and obtained in various numerical calculations.
The second independent parameter $k_0$ should
be varied in the region of $1 \ GeV$ - the characteristic
scale of low energy QCD.
  Varying now independently this pair
of parameters $m_d$ and $k_0$
numerically, one can calculate all chiral QCD parameters using
the above derived formulae (6.1-6.5).
 
 From the value of the pion decay constant in the chiral
limit (5.40) as well as from the range selected first for $m_d$
(6.17) and on account of
(6.1) and (6.2), it follows that the momentum $k_0$ should always
satisfy the upper and lower boundary value conditions, namely
\begin{equation}
691.32 \le k_0 \le 742.68 \ (MeV).
\end{equation}
See Fig. 3. In the ranges determined by (6.17) and (6.18) the
vacuum
energy density of the nonperturbative gluons (6.5) changes its
sign (Fig. 6) and becomes positive which is undesirable.
It is easy to show that this is a result of that that the lower
bound
chosen for the dynamical generated quark mass in (6.17) is too
low. Indeed, the vacuum energy density (6.5) vanishes
at the critical point $z^{cr}_0 = 1.45076$ (Fig. 8). Then from
(6.2), calculated at this point, it follows that
\begin{equation}
 k_0 \leq 2.26 m_d.
\end{equation}
Because of this inequality the vacuum energy density (6.5)
will always be negative as it should be and only at the critical
value
\begin{equation}
 k^{cr}_0 = 2.26 m_d
\end{equation}
becomes zero (worst case). From the chosen interval for $m_d$
(6.17) and the obtained interval for $k_0$ (6.18) it follows
that the ratio between the corresponding lower bounds $k_0/m_d =
691.32/300 = 2.3044$ does not satisfy the relation (6.19) while
this ratio for the corresponding upper bounds $k_0/m_d =
742.68/400 = 1.8567$ does satisfy it.
This explicitly shows that the lower
bound for $m_d$ in (6.17) was incorrectly chosen. It is, as
mentioned above, too low. The exact lower bound for
$m_d$ can be found from the relation (6.20) as
\begin{equation}
 742.68 = 2.26 m_d,
\end{equation}
and (6.17) becomes
\begin{equation}
 328.62 \le m_d \le 400  \ (MeV).
\end{equation}
In the ranges determined by (6.18) and (6.22), the vacuum energy
density (6.5) will be always negative because any combination
(ratio) of $k_0$ and $m_d$ from these intervals will satisfy the
relation (6.19). However this is not the whole story. A new lower
bound for $m_d$ leads to a new lower bound for $k_0$ as well.
Combining now this new lower bound (6.22) with the
chiral value of the pion decay constant (5.40), one obtains a new
lower bound for $k_0$ as well.
 
  As we emphasized several times, $k_0$ should be treated
as a momentum which separates the nonperturbative phase (region)
from the perturbative one (region), so that in a new region
obtained
for $k_0$ the nonperturbative effects, such as quark confinement
and DCSB, begin to play a dominant role. This is the region
where confinement occurs. From now on, let
us call this scale for $k_0$ a confinement scale (in the
chiral limit) and denote it $\Lambda_c$. So the final numerical
limits for the confinement scale are
\begin{equation}
707 \le \Lambda_c \le 742.68 \ (MeV).
\end{equation}
In the intervals determined by (6.22) and (6.23) the vacuum energy
density due to the nonperturbative gluon contributions,
$\epsilon_g$, will be always negative (see Fig. 9).
 
   It is important
that any value from the obtained interval
(6.23) is allowed but not each combination
of $\Lambda_c$ from interval (6.23) and
$m_d$ from interval (6.22) will automatically yield the
value of the pion decay constant given by (5.40). Therefore
it is necessary to adjust values of $m_d$ from (6.22) and
$\Lambda_c$ from interval (6.23), and vice
versa, as it is seen in Fig. 10.
This means that $m_d$ is in a close relationship with $\Lambda_c$.
Moreover, completing the above mentioned adjusting procedure, one
finds that $\Lambda_c$ is nearly double the generated quark
mass $m_d$, i. e.
\begin{equation}
\Lambda_c \approx  2 m_d.
\end{equation}
This confirms that $\Lambda_c$ and $\Lambda_{CSBq}$, defined by
(5.15) or (6.6), are nearly the same indeed.
In the previous calculation
scheme the adjusting procedure was automatically fulfilled because
of relation (6.7). There is a
close relationship between  $\Lambda_{CSBq}$ and $\Lambda_c$ on
one hand, and the double generated quark mass $m_d$, on the
other.
 
  Evidently interval (6.23) for $\Lambda_c$,
along with the new range for $m_d$ (6.22), uniquely determines
the upper and lower bounds for all other chiral QCD
parameters considered here. Just like in the previous case, our
numerical results are shown in Table 2, where the shorthand
notation
${\langle{0}|G^2|{0}\rangle}$ stands for the gluon condensate
${\langle{0}|{\alpha_s \over
\pi}G^a_{\mu\nu}G^a_{\mu\nu}|{0}\rangle}$. Our numerical bounds
for the vacuum energy density $\epsilon$ need some additional
remarks. We emphasize that bounds for $\epsilon$
is not the sum of bounds for $\epsilon_q$  and $\epsilon_g$.
Indeed, the upper and lower bounds for $\epsilon_q$
are achieved at the upper and lower bounds for $m_d$
($\Lambda_c$)
while for $\epsilon_g$ they are achieved at the lower and
upper bounds of $m_d$ ($\Lambda_c$). At the same time, the
intervals for $\epsilon$ differ very little from each other in
either case (see Table 2).
 
  Let us now prove the relation (6.24). We have already learnt
that
correct values of $k_0$ belongs to the interval for $\Lambda_c$
(6.23). Then identifying $k_0$ with $\Lambda_c$ in (6.19), one
obtains
\begin{equation}
\Lambda_c \leq (2 + 0.26) m_d = \Lambda_{CSBq} +0.26 m_d,
\end{equation}
so
\begin{equation}
\Delta = \pm (-1 + {\Lambda_c \over \Lambda_{CSBq}}) \leq 0.13,
\end{equation}
where the positive sign corresponds to $\Lambda_c \geq
\Lambda_{CSBq}$
and the negative one holds when $\Lambda_c \leq \Lambda_{CSBq}$.
In derivation of both these relations we used definition (6.6).
The maximum deviation will be achieved only at the
critical point, i.e. when the contribution to the vacuum energy
density
of the nonperturbative gluons vanishes. Of course, this is not
the case and these two scales are very close indeed.
 
  It is worth underlining once more that besides
good numerical results obtained
in this subsection (see Table 2), we have established the
existence of a realistic lower bound for the dynamically generated
quark masses. In each calculated case their numerical values are
shown in
Table 2. Thus one concludes that the vacuum energy density of the
nonperturbative gluons is sensitive to the lower bound for $m_d$.
The second important result is that we have
clearly shown that the confinement scale $\Lambda_c$ and DCSB
scale $\Lambda_{CSBq}$ are nearly the same indeed.

\section{Conclusions}

 In the theoretical part of our paper we have seen that the
ZME model of both the quark confinement
and DCSB reveals several desirable and promising features.
The quark propagator has indeed no poles (quark confinement
theorem). A single quark (heavy or light) is already off
mass-shell by virtue of the ZME effect in the
ground state of QCD - true vacuum. Let us recall that a "single"
quark is understood by means of a propagation described by the
solution to the quark SD equation.  Our model also implies DCSB
at the
fundamental quark level. The system of equations (3.29-3.30)
for the ground solution (3.32-3.33) of the quark propagator (2.2)
$forbids$ a chiral symmetry preserving solution
\underline{($m_0 = B(t) = 0, \  A(t) \ne 0 $).}
Moreover, a chiral symmetry violating solution
\underline{($m_0 = 0,\ B(t) \ne 0, \ A(t) \ne 0 $)} for the quark
SD equation $is \ required$. Thus chiral symmetry is certainly
dynamically (spontaneously) broken in QCD because of the ZME
effect. The finite
corrections, coming from the intermediate and UV regions, to the
solution for the quark propagator, because of the much less
singular behaviour in the IR, cannot destroy these features of
the
ground solution. The absence of the pole singularities in the
quark
propagator (quark confinement) and DCSB at the fundamental
quark level are closely related.
The intrinsic link between these two important
phenomena is a direct consequence of the nonperturbative,
IR singular structure of the QCD ground state - true vacuum.
Thus the ZME effect alone provides the dynamical mechanism for
both DCSB and quark confinement phenomena in QCD.
 
From our theoretical considerations, which were numerically
confirmed in the second part of our
paper, it follows that the formation of the
dynamical quark, as well as the quark and gluon condensates,
at a certain scale becomes
inevitable if the zero modes are indeed enhanced in true vacuum of
QCD. Dynamical quarks must be formed and this should be
accompanied by the formation of quark and gluon condensates.
The formation of quark and gluon condensates necessarily
leads to the dynamical quarks.
Moreover the existence of the dynamical quarks is the first
necessary step for the pion to be a Goldstone state. Its
decay constant, in the chiral limit, Eq. (5.9), depends on the
mass of the dynamical quark. The Goldstone nature of the pion was
also independently confirmed by the investigation of the
structure of the dynamical singularities of the corresponding
axial-vector WT identity in the chiral limit (see Section 2 and
our paper [19]).
 
Thus the dynamical
quark, defined as the inverse of the quark propagator at zero
point (5.11), is one of the key objects of our model [74]. Our
numerical
results clearly show that its definition in the chiral limit
(5.11) is in fair agreement with the definition of a scale
responsible for DCSB at the quark level (5.15) and consequently
the confinement scale as well. One may conclude that the
nonperturbative ansatz (5.6), which allows to take into
account nonperturbative contributions to the pion decay constant
coming from the second diagram in Fig. 2, works in the right way.
The vacuum energy density calculated through
the effective potential for composite operators, on one hand,
numerically is in a self-consistent relation with all other
chiral QCD parameters. On the other hand, it agrees very well with
values provided by QCD sum rules and the RILM
(see, however, the next section). Moreover we
find that contributions to the
vacuum energy density, coming from the confining
quarks with dynamically generated masses, $\epsilon_q$, and the
nonperturbative gluons, $\epsilon_q$, are nearly the same indeed.
 
 As it was previously noted, the QCD coupling constant, in the
massless quark case, plays no independent role. We remind
the reader that the ZME effect leads
to the strong regime for the effective (running) coupling
constant (item III of the Section 3). For this reason, and
because of the above mentioned dimensional
transmutation phenomenon [62], analysis, based on the existence
of a critical value of the strong coupling
constant (at which DCSB and quark confinement occur), should be
replaced by the analysis based on the existence of a certain
scale at which these phenomena become essential.
This was one of the highlights in our paper.
 
  Let us emphasize independently that the
numerical investigation of the chiral sector in QCD, in our
calculation schemes, shows clearly that the scales of both
nonperturbative phenomena (quark confinement and DCSB) are nearly
the same, i.e.
\begin{equation}
\Lambda_c \approx \Lambda_{CSBq} = 2 m_d.
\end{equation}
The scale (7.1) has a clear physical meaning in our model.
It determines a scale at which transition between nonperturbative
and perturbative phases occurs.
The definition (7.1) is relevant in the case of the light quarks
only ($u$, $d$ and $s$).
Let us note also that in our paper [20], on the
basis of naive counting arguments, it was explained why the
scale $\Lambda_{CSBq}$ should be approximately the half of the
corresponding scale of DCSB at the hadronic level $\Lambda_{\chi}$
- the scale of effective field theory (CHPTh) [28, 52, 53].
 
  For reasonable ranges obtained for $m_d$ (see Tables 1 and
2) and for the choice (6.7), one certainly comes to
the self-consistent numerical picture of the nonperturbative
dynamics in
our general approach to QCD at large distances. Our model
depends only on two independent (free) parameters
which have a clear physical sense
and therefore it provides a well-controlled calculation scheme
(in various forms). In principle, we are able to numerically
calculate integrals in both schemes to any requested accuracy.
It is also worthwhile to compare the relative simplicity of our
calculations with the indisputable complexity of the lattice ones.
The proposed dynamical
quark propagator approach complemented by the CHPTq can be
applied with the same success for investigating analytically and
computing numerically from first principles any other low energy
physical quantities in a self-consistent way.

\section{Discussion}

  In the preceding two sections we have obtained and summarized
our results which are direct
(quantitative) consequences of the zero modes enhancement (ZME)
effect in the framework of our
general approach to low energy QCD. Until now we have not
made any attempts to interpret the above mentioned effect from
the viewpoint of dynamics.
 Let us make now a few detailed remarks of semi-intuitive, as well
as semi-speculative, character which, will, nevertheless, shed
some light
on our understanding of the actual dynamical mechanism of quark
confinement and DCSB might be interpreted with the help of the ZME
effect. People who do not like this way of thinking are
recommended
to skip this section. Still, we think, this line of thought
illustrates the
rich possibilities of this effect in QCD. Moreover, the
section contains
predictions for more realistic values of the vacuum energy density
(the bag constant, apart from the sign)
and the gluon condensate.

\subsection{A possible dynamical picture of quark confinement and
        DCSB within the ZME effect in QCD }

  "Once upon a time in America", Susskind and Kogut [39] have
noticed that "the absence of quarks and other colored objects
could only be understood in terms of an IR divergences in the
self-energy of a color bearing object". In our
approach gluons remain massless, only zero modes are enhanced.
We will discuss in more detail contributions
to the self-energy of the colored quark leading first to the
dynamical and then to constituent
quarks in the context of the ZME which is caused by the
nonperturbative
IR divergences in the true vacuum of QCD. In other words, we will
try to explain below that the ZME model of quark
confinement and DCSB is a direct realization and development of
Susskind's and Kogut's program.
 
In order to clarify the dynamical picture which may lie at the
heart of our model, let us introduce, following
Mandelstam [37], two sorts of gluons. The actual (external) gluons
are those
which are emitted by a quark and absorbed by another one, while
the virtual (internal) gluons are those which are emitted and
absorbed by
the same quark. At first sight  this separation seems to
be a simple convention but we will show below that it has a firm
dynamical ground, thus it makes our understanding of the above
mentioned picture more transparent.
 
 Let us consider now all the possible contributions to
the self-energy of a single quark. The most simplest one is
shown in Fig. 11. Let us recall that the same self-energy
diagram occurs also in the quantum electrodynamics (QED). In
contrast to QED, there is an infinite number of additional
contributions to the self-energy of a single quark because of
the non-abelian nature of QCD, i.e. because of the direct
interaction between virtual gluons which is absent in QED. Some
of these are shown
in Fig. 12. So, from the point of view of the contributions to the
self-energy of a single quark, the zero modes are indeed enhanced
in QCD in comparison with the electron self-energy in QED.
The self-interaction of virtual gluons alone removes
a single quark from the mass shell, making it an effective
(dynamical in the chiral limit) object. This was the context of
the quark confinement theorem described in Section 3.
The effective mass of this object is defined as the inverse
of the full quark propagator at zero point (Eq. (5.13)).
 
 But this is not the whole story yet in QCD because up to now we
took
into account only contributions induced by the virtual gluons
alone.
The actual gluons emitted by one quark can contribute to the
self-energy of another quark and vice versa. The simplest diagrams
of this process are shown in Fig. 13. Moreover
contributions shown in Fig. 14
are also inevitable and they describe the process of
the convertation (transformation) of virtual gluons into
actual ones and the other way around.
Thus we consider diagrams, of these type, not as
corrections to the cubic and quartic gluon vertices but rather as
additional contributions to the self-energy of the quarks.
Contributions to the self-energy of each quark will be essentially
enhanced in the presence of another quark. In other words
each quark additionally enhances the interaction with the
vacuum (zero modes) of another quark.
It seems to us that the IR strong singular asymptotics of the full
gluon propagator (3.1) or equivalently (3.8) effectively
describes this phenomenon in QCD correctly, i.e enhancement of the
zero modes by virtue of self-interaction of virtual (internal) and
actual (external) gluons.
 
It is quite plausible that, at large
distances between quarks, actual gluons emitted by each quark do
repeatedly succeed to convert into virtual ones and vice
versa. This leads to a multiple enhancement of
the zero modes of each quark. We think that these
additional contributions to the self-energy of each effective
quark makes it a constituent object in our model. The mass of
the constituent (heavy or light) quark becomes the sum of three
terms, namely
\begin{equation}
 m_q = m_{eff} + \Delta = m_0 + m_d + \Delta,
\end{equation}
where we used definition (5.13). All terms on the right
hand side have clear physical sense.
The first term is, obviously,  the current
mass of a single quark. The second one $m_d$ describes
contributions to the constituent quark mass induced by the
self-interaction of virtual gluons alone, while the third term,
$\Delta$, describes contributions to the constituent quark mass
which come from the process of the convertation of
actual gluons into virtual ones as it was discussed above.
This way our model provides a natural dynamical foundation of the
current-effective (dynamical)-constituent transformation [75] of
the quark degrees of freedom on the basis of non-abelian
character of the gluon fields. The existence of a nonzero
$\Delta$ is principal for our model but numerically it should not
be large, even for light quarks. Our intuition (based on the
obtained numerical results) tells us that it is only
of the order of a few per cent of the displayed values of
$m_d$ (Tables 1 and 2).
 
Remember that this naive formula shows
only the nature of the contributions to the constituent quark
mass. The contributions are so mixed up in reality that they
can not be separated from each other. There exists
an infinite number of
possible, topologically complicated, configurations of the vacuum
fluctuations of the non-abelian gauge (gluon) fields contributing
to the self-energy of each quark while making them constituent
objects. The true vacuum of QCD, however, is not settled by
these fluctuations alone, its structure is much
more richer [70] than that (see also the discussion below).
 
  Certainly, a finite number of favorable, topologically distinct
vacuum configurations, which
minimize the energy of the bound states, should exist. It is
hard to believe that in the real world of four dimensions, the
favorable topologically complicated configurations are strings
or planar ones. In this context, it is important to
comprehend that the linearly rising quark-antiquark potential,
at large distances, nicely showed by recent lattice calculations
[76], is not a privilege of the planar or string configurations
only. Though the above mentioned linear potential does not
contradict the ZME effect, nevertheless the
potential concept in general is a great simplification of
the real dynamical picture which emerges from our model.
As it was mentioned in Section 3, the enhancement of zero
modes necessary leads
to full vertices while the potential concept of the CQM
[32, 33, 61, 64] is based on point-like ones.
 
   One may say that at large distances interaction
between quarks proceeds mainly through the vacuum and thus leads
first to the formation of the constituent quarks and then to the
formation of bound states of these constituent quarks. According
to the ZME effect
this process should be accompanied by the condensation of the
quark and gluon loops - quark and gluon condensates.
Hadrons are bound-states of the constituent
quarks strongly interacting with the vacuum of QCD at the expense
of each other. At the same time, direct interaction
(through the actual gluons) between constituent colored objects
becomes weeker. Also, they should be surrounded by quark
condensates playing the role of some external
field (see Subsection C below). From our dynamical
picture it follows that hadrons are composed of heavy
objects (constituent quarks) because of the
enhanced contributions to their self-energy, and because of this,
the quarks interact weakly. The reason
of the relatively week interaction between the constituent quarks
is that the biggest part of the actual gluons is expected, at
large distances,
to be converted into virtual ones which contribute to the
self-energy of each quark.
 
There exists an interaction between electrons
and positrons in QED which
goes through the vacuum in the sense described
above, but it is strongly suppressed because of the small
magnitude of the coupling constant (fine structure
constant) and the absence of the direct interaction between
photons. Consequently, there does not
exist such an object as the "constituent electron" in QED. In
other words the perturbative vacuum of QED can produce only small
corrections to the self-energy of the electron leading to a
"clothed or dressed" electron, while the nonperturbative vacuum
of QCD produce not only corrections but substantial contributions
to the self-energy of the quarks which make
them constituent objects. Thus a single electron can be in a free
state and remain "clothed"
but a constituent quark can not be in a free state as a "clothed"
electron can because it becomes constituent in the presence,
at least, of another quark which itself becomes constituent too.
 
It is plausible that
the energetically advantageous configurations of
vacuum fluctuations (leading to the formation of the
bound-state of the constituent quarks) occur at
a certain distance between constituent quarks. This
configuration will be completely deformed (or even destroyed) if
one attempts to
separate these constituent quarks further from each other.
The nonperturbative vacuum of QCD is filled with
quark-antiquark virtual pairs which consist of various
components of quark degrees of freedom (light, heavy,
constituent, dynamical, etc).
This is an inevitable consequence of the ZME effect.
As a result of the above mentioned nontrivial topological
deformation, at least one quark loop will be certainly "cut" (see
also Subsection C below).
We may, for convenience, think of this as such a topological
deformation which allows for quarks from
the loop to recombine with the initial constituent quarks. It is
evident that the breaking of the gluon line is not so important as
the above mentioned cut of the quark loop which is
equivalent to the creation of the corresponding
quark-antiquark pairs from the vacuum. The vacuum of QCD
will be immediately rearranged and, instead of
"free" constituent quarks, new hadron states will occur.
 
   Here let us remind the reader that the
constituent quark can
not go on mass-shell (quark confinement theorem, Section
3). Apparently, at zero temperature to confine such quarks in
order to generate
new hadrons states is much more energetically profitable
than to keep them always "free" and off mass-shell.
At nonzero temperature the transition from hadron matter to
quark-gluon plasma is possible [73, 77]. In both cases, at
zero temperature (confining quarks inside hadrons) and at nonzero
temperature ("free" in quark-gluon plasma) quarks always remain
off mass-shell. In other words, in the ZME model of the
vacuum of QCD, deconfinement should be interpreted as the
screening of the confining potential and the dissociation of
bound states; nevertheless, quarks can never go on
mass-shell. This clearly shows once again the restricted sense of
the potential concept.
Thus our model admits the existence of the
quark-gluon plasma because, by melting
more and more hadron matter, it is impossible to
remove quarks from off mass-shell.
 
  At short distances the situation is completely different from
the above described. Indeed, in this case the actual gluons
emitted by each quark do not repeatedly succeed to convert into
virtual ones. So, at these distances, interaction between
quarks proceeds mainly through the exchange of actual gluons.
This means that the interaction of the constituent quarks with the
vacuum, due to the above mentioned process of
convertation, is essentially decreased. So they become valence
quarks. The intensity of the process of convertation determines
the constituent-valence transformation. One can say that
the constituent (valence) quarks are " valence
(constituent)-in-being quarks". At low energies a valence quark
becomes constituent by absorbing self-interacting gluons and
sea quarks with the help of another constituent quark. At high
energies
these degrees of freedom will be lost and all constituent
quarks become valence ones.
 
 The potential concept (based on point-like vertices) becomes
relevant in this case. At
the experimental large momentum scale, the virtual photon sees the
hadron as made of valence quarks, sea quarks and gluons.
The Coulomb-type behaviour is a good approximation for the
interaction
between quarks which, at this scale, become rather simple objects
(almost massless).
Contrary to the small momentum scale, at which complicated
objects are quarks, at this scale gluons themselves
play a major role. At large distances quarks
interact strongly with the vacuum, and therefore interaction
between
them becomes week, while at short distances gluons interact
strongly with the vacuum which produces a strong anti-screening
effect in it, leading to asymptotic freedom [17].

\subsection{The Okubo-Zweig-Iizuka selection rule}

In the context of the ZME
model of quark confinement and DCSB it becomes
clear that the topological rearrangement of the vacuum by means
of the $direct$ annihilation of the initial (final)
constituent quarks, entering the same hadrons, is hardly
believable. In fact, what does the above mentioned $direct$
annihilation mean? This would mean
that the initial (final) constituent
quarks, emitting (absorbing) a number of
gluons, can annihilate with each other without the break-up of
the corresponding quark loops in our model. The initial (final)
constituent quarks always emit and adsorb
gluons in each preferable configuration.  Nothing interesting
should happen during these
processes. This is a normal phase of each preferable
configuration
and it describes only its trivial rearrangement. Any nontrivial
rearrangement of the vacuum can only begin with
cutting the quark loop. As it was mentioned above, this is
equivalent
to the creation of a quark-antiquark pair from the vacuum. Then
the annihilation of the initial (final) constituent quarks with
the corresponding quarks, liberated from the loop, becomes
possible. Diagramatically this looks like a $direct$ annihilation
(see Fig. 15).
The probability to create the necessary pair, in order to
annihilate the
initial (final) constituent quarks, is rather small, so, in
general, the annihilation channel must be suppressed.
It is much more
probable for the quarks liberated from the loop to recombine
with the initial (final) constituent quarks in order to generate
new hadron states.
In more complicated cases (when many quark loops are cut) the
annihilation of the initial (final) constituent quarks becomes
more
probable and this process will compete with the process of the
recombination of the initial (final) constituent and liberated-
from-the-vacuum quarks to generate new hadron states directly.
Ending these general remarks, first let us analyse
 mesons consisting of heavy quarks, for example $c \bar{c}$
systems.
 
 The heavy
constituent quarks inside hadrons are much closer to each other
(the distance between them is of the order $m^{-1}_h \ll
m^{-1}_q$,
where $m_h$ and $m_q$ denote the masses of the heavy and light
constituent quarks, respectively) than their light counterparts.
At first sight this should promote the annihilation of the
initial (final) constituent quarks. However at short distances
the interaction between them proceeds mainly not through the
vacuum
fluctuations but via the exhange of the actual gluons as this was
explained above. This means, in turn, that the number of virtual
loops which should be cut is small. Also the
probability to cut heavy quark loop, or equivalently to create a
heavy quark-antiquark pair from the vacuum, is much less
than to create, for example, a light pair. Apparently,
the fluctuations
in the density of instantons (see next section) and condensates
during the vacuum's rearrangement also come into play.
Thus the process of the rearrangement
of the vacuum, leading to the transition between hadrons (their
strong decays) on the basis of the annihilation of the initial
(final) constituent quarks, as shown in Fig. 15, should be
strongly
suppressed in comparison with the process of the recombination
of the initial (final) constituent quarks with the quarks
liberated from the vacuum. This is shown in Fig. 16. Thus one
comes to a dynamical explanation of the
famous Okubo-Zweig-Iizuka (OZI) selection rule [78].
In the case of heavy quarks it is in agreement
with the standard explanation of
the OZI rule which states that the
QCD coupling constant becomes weak at short distances and
suppresses
the annihilation channel. However this argument fails to explain
why the violation of the OZI rule for the pseudoscalar octet is
bigger than for the vector one.
 
   Let us analyse this problem in our approach.
Light constituent quarks inside pseudoscalar and vector mesons
are at relatively large distances ($\sim m^{-1}_q$ from each
other)
than their heavy counterparts, for example in $c \bar{c}$ systems.
This means that interaction between them is mediated mainly by the
vacuum fluctuations which provide plenty  of various
quark loops to be cut during the process of the vacuum
rearrangement. So the annihilation channel should not be strongly
suppressed for these octets. Indeed, the violation of the OZI rule
in the pseudoscalar mesons is not small, but for the vector mesons
it is again small, i.e.  comparible to the violation in the
$c \bar{c}$ systems. Our model
provides the following explanation for this problem:
    The same quark-antiquark pair in pseudoscalar and vector
mesons is in the same $S$-state. The only difference between
them is in the relative orientation of the quark spins.
Quark and antiquark spins are oriented
in the same direction in the vector mesons, while in the
pseudoscalar mesons their
orientation is opposite. This is schematically shown in Figs. 17
and 18.
For light constituent quarks spin effects become important, while
for heavy constituent ones such a relativistic effect as spin and,
in particular, its orientation is not so important. As it was
repeatedly
mentioned above, a nontrivial rearrangement of the vacuum in
our model always
starts from the cut of the quark loops. In order to analyse
the violation of the OZI rule, from the point of view of
annihilation of spin degrees of freedom, let us think
of quark loops as "spin loops". In pseudoscalar mesons at
least one spin-antispin liberated-from-the-vacuum-pair is needed
to annihilate the initial pair. This is schematically shown in
Fig. 17.
In vector mesons at least two spin-antispin
pairs are needed for this purpose, but, in addition, an
intermediate meson (exited) state certainly appears, see Fig. 18.
This means that the annihilation channel for the vector mesons,
unlike for the pseudoscalars ones, is suppressed.
It is worth noting that the OZI rule is a selection rule and it is
not a conservation law of some quantum number. So its breakdown is
always possible and suppressed processes may proceeds through the
appropriate intermediate states [77]. Exactly this is
shown in Fig. 18 schematically . All
this explains the violation of the OZI rule in the
pseudoscalar channel in comparison with the vector one in our
model.
For vectors mesons the decays $\phi \rightarrow 3 \pi,
\rho \pi$, proceeding
through the annihilation channels, are suppressed in comparison
with the decay $\phi \rightarrow K^+K^-$ which occurs via the
recombination channel; while for pseudoscalar mesons, say,
the decay $\eta \rightarrow 3 \pi$ is not suppressed [54].
 
To conclude this topic, let us underline that our model
of the QCD vacuum on the basis of the ZME effect provides an
explanation of the OZI rule on a general ground for all mesons
($q \bar{q}$ systems), in particular, for $c \bar{c}$,
pseudoscalar,
vector, etc ones. In order to confirm our qualitative explanation
of the OZI selection rule quantitatively, it is necessary to
calculate the strong decay widths of mesons in our model, which
is beyond the scope of the present paper and will be done
elsewere.

\subsection{Instantons}

 The main ingredients of the QCD vacuum, in our model, are
quark and gluon condensates, quark-antiquark virtual pairs (sea
quarks) and self-interacting nonperturbative
gluons. The vacuum of QCD has, of course, much more remarkable
(richer) topological structure than this. It is a very
complicated medium and its topological complexity means that its
structure can be organized at various levels and it
can contain perhaps many other components [1, 70] besides
the above mentioned. There are a few models of the nonperturbative
vacuum of QCD which are suggestive of what a possible confinement
mechanism might be like (see recent paper [80] and references
therein).
We will not discuss these models; let us only mention that
the monopole vacuum of Mandelstam and  't Hooft as well as the
mechanism of the confining medium recently proposed by Narnhofer
and Thirring [81] also invokes the
ZME effect [37, 80, 82]. Let us ask one of the main questions now.
 
 What is a mechanism like which initiates a topologically
nontrivial rearrangement of the vacuum?  It is
already known, within our model, that this may begin with the cut,
at least, of one
quark loop and therefore, at least, one quark-antiquark pair
emerges from the vacuum. Why can the quark loop be cut at all
and what prevents the quarks from the cut loop to
annihilate again with each other? The fluctuations in the
nonperturbative vacuum of QCD must
exist which do the above mentioned job.
This is an inevitable consequence of our model of the vacuum.
We see only one candidate for this role in four dimensional QCD,
namely instantons and anti-instantons and their interactions
[83]. In this sense let us discuss the
possible role of instantons in our picture of quark confinement.
At this stage of our knowledge of the detailed mechanism of
confinement, one may only make a few remarks which will be
necessarily qualitative.
 
Instantons are classical (Euclidean) solutions to the
dynamical equation of motion of the nonabelian gluon fields
and represent topologically nontrivial fluctuations of these
fields [1, 83]. Self-interaction of gluons should be important
for the existence of the instanton-like fluctuations
even at classical level.
In the above mentioned RILM [69], light quarks
can propagate over large distances in the QCD vacuum by simply
jumping from one instanton to the next one. In contrast, in
our model the propagation of all quarks is determined by the
corresponding SD equations (due to the ZME effect) so that they
always remain off mass-shell. Thus we do not need the
picture of jumping quarks. As opposed to the
RILM, we think that the main role of the instanton-like
fluctuations is precisely to prevent the quarks and gluons from
freely propagating in the QCD vacuum. Running against
instanton-like fluctuations, the quarks undergo difficulties in
their propagation in the QCD vacuum which is a very
complicated inhomogenious medium. Along with
the instanton component, it consists of a confining quark
component, the nonperturbative gluon component, etc. As we have
already learned the confining quark component, in turn, consists
of
various quark degrees of freedom such as dynamical, light, heavy,
constituent ones, etc. At some critical value of the instanton
density the
free propagation of the virtual quarks from the loops, apparently,
become impossible so they never annihilate again with
each other. Obviously, this is equivalent to the creation of the
quark-antiquark pairs from the vacuum. From this
moment the nontrivial rearrangement of the vacuum may start.
The role
of the instanton-like fluctuations appears to be "cutting"
the quarks loops and preventing them from the immediate
annihilation of quarks and antiquarks liberated from the loops.
Their main task is to promote transitions between hadrons,
i.e. in our
terminology they destabilize energetically
advantageous (dominant) configurations of the vacuum fluctuations
which lead to hadron states. One of the main features of the
instanton-induced effects is tunneling between topologically
distict vacuums in Minkovski space [1, 84]. It seems to us that
our understanding of their role in the QCD vacuum structure
does not contradict this.
 
Being classical (not quantum!) fluctuations, instantons
can cut the quark loops in any points even in the quark-
gluon vertices. A simple cut of the quark loops, as it was
repeatedly emphasized above, is equivalent to the creation of the
corresponding quark-antiquark pairs from the vacuum. Let
us ask now what happens if all the external (actual) gluon lines
will be cut from the quark loops by the instantons. Well, in this
case each quark loop becomes a closed system. Because of the
vacuum pressure they immediately collapse and one obtains nothing
else but quark condensate if, of course, all internal gluon lines
can
be included into the quark self-energy. Another scenario is also
possible when not all the internal gluon lines can be included
into
the quark or gluon self-energy (see, for example, the two-particle
irreducible (2PI) vacuum diagrams, sketched in Fig. 19).
In general, the presence of the
internal gluon lines in the vacuum diagrams prevent them from
collapsing
(because they counterbalance the vacuum pressure) and consequently
they should contribute to the vacuum energy density.
 
It would be suggestive to conclude that the same mechanism
works in order to produce gluon condensates despite the much more
complicated character of the gluon self-interactions. But this
is not the case indeed, since there is a principal difference
between quark and gluon condensates. The former ones do not
contribute to the vacuum energy density and, in this sense, they
play the role of some external field. While the latter ones, as
was
shown first in Ref. 49 and numerically confirmed in our paper, are
closely related to the vacuum energy density. Like we noted
previously, nothing interesting should be happen if the instantons
cut
the gluon line in one point. Let us imagine that many gluon lines
will be cut by the instantons in many points. The vacuum
becomes filled up with gluon pieces (segments) which, due to
the existence
of the gluon strong self coupling, can recombine, in principle,
as some colour singlet bound-states --
gluonia or glueballs. Unlike quark condensates,
glueballs should have internal pressure because of the strong
self-interaction of composite gluon segments which prevents them
from collapsing. In turn, this means that the glueballs should be
heavy enough. At  this stage we see no reason why the ZME model
of the QCD vacuum should forbid non-quark-antiquark bound-states,
such as glueballs, multiquark states or even hybrids
($q \bar{q}g$) [54].
 
  Our numerical results were obtained in the chiral limit but our
qualitative discussion was a general one. We are mainly concerned
with realistic (nonchiral) QCD. Let us now briefly discuss the
important case of massless quarks (i. e. the chiral limit). The
pseudoscalar
mesons (consisting of light quarks) are Goldstone states so their
masses remain zero in the chiral limit even in the presence of
DCSB. From our model it follows that the existence of the
instanton-type fluctuations in the true vacuum promote strong
meson decays by preventing quarks and gluons from freely
propagating in it. One can conclude in that
instanton-type fluctuations should be totally suppressed in this
case in
order to provide stability for the massless Goldstone states since
massless particles cannot decay. This feature of the
instanton physics in the massless quark case was discovered
by t' Hooft [71]. In the presence of DCSB, however, the
instanton-type fluctuation are restored [70, 71] but, apparently,
the contribution of the instanton component to the vacuum energy
density still remains small. So the dilute gas approximation for
the
instanton component seems to be relevant in this case. Not going
into details of the instanton physics (well described by Callan,
Dashen and Gross in Ref. 70), let us only emphasize that at short
distances the density of small size instantons should  rapidly
decrease and conversely increase at large distances where large
scale instantons, anti-instantons and their interactions also come
into play. Otherwise it would be
difficult to understand the role which we would like to assign
to the instantons in our model. This is in agreement
with the behaviour of the instanton component of the QCD
vacuum at short and large distances described in the above
mentioned paper [70] as well as
with our understanding of the actual dynamical mechanism of quark
confinement and DCSB described in this section.
 
The nontrivial rearrangement of the vacuum
can start only when the density of instantons achieves
some critical values, different for all distinct vacuums.
For this reason, despite being a classical phenomena,
instantons should nevertheless
contribute to the vacuum energy density through the above
mentioned quantum tunneling effect which is known to lower the
energy of the ground-state. We have already calculated the
contributions to the vacuum energy density of confining quarks
with dynamically generated masses and nonperturbative gluons
(Tables 1 and 2). So the total vacuum energy becomes (as
minimum) the sum of three quantities now, namely
\begin{equation}
\epsilon_t = \epsilon_I + \epsilon_g  + N_f \epsilon_q
\end{equation}
where  $\epsilon_I$ describes the instanton component of the
vacuum. We introduce also the explicit dependence on the number
of different quark flavors $N_f$ since $\epsilon_q$ itself is
the contribution of a single confining quark. Of course, this
should be valid for the non-chiral
(realistic) case as well. The distinction will be in the actual
values of each component, apart from, maybe, $\epsilon_g$.
 
  Up to now our discussion of a possible role of the
instanton fluctuations in our model was quite qualitative.
Let us now run a risk and make a few quantitative predictions.
As it was noticed, $\epsilon_I$
should be small in the chiral limit and in the presence of DCSB,
i. e. for light quarks with dynamically generated masses. However,
the same conclusion seems to be valid for heavy quarks as well in
our model. Heavy quarks are at short distances from each other (in
mesons) at which
the nonperturbative effects, such as instantons and enhancement of
zero modes, are suppressed. Thus the dilute gas approximation for
the instanton component seems to be applicable to light quarks
with dynamically generated masses as well as to heavy quarks.
In other words, $\epsilon_I$ for light quarks with dynamically
generated masses and for heavy quarks is nearly the
same in our model.
If this is so then it is worth assuming, following the authors of
Ref. 49, that the light and heavy quarks match smoothly. This
allows one to choose the average value between (6.15) and (6.16)
for the instanton component of the vacuum
energy density $\epsilon_I$ in (8.2), i.e. namely
$\epsilon_I \simeq 0.0034 \ GeV^4$. Then our
predictions for the total vacuum energy $\epsilon_t$ (in the
case of $N_f$ light confining quarks with dynamically generated
masses)
and the corresponding values of the gluon condensate are listed in
Tables 3 and 4, 5 for both calculation schemes A and B,
respectively.
 
  It is worth reproducing explicitly some interesting particular
values of the total vacuum energy density and the corresponding
values of the gluon condensate. Thus for a pure gluodynamics
($N_f=0$) one has
\begin{eqnarray}
\epsilon_t &\simeq& - 0.005 \ GeV^4, \nonumber\\
\langle{0}|{\alpha_s \over
\pi}G^a_{\mu\nu}G^a_{\mu\nu}|{0}\rangle &\simeq& 0.0177 \ GeV^4,
\end{eqnarray}
and
\begin{eqnarray}
-0.00661 &\leq \epsilon_t \leq& - 0.003837 \ (GeV^4), \nonumber\\
0.0136 \leq &\langle{0}|{\alpha_s
\over \pi}G^a_{\mu\nu}G^a_{\mu\nu}|{0}\rangle& \leq 0.0235 \ (GeV^4).
\end{eqnarray}
Here and below the numbers correspond to the approximation of the
pion decay constant in the chiral limit by its standard value.
 
  For the most realistic case $N_f=2$ one obtains
\begin{eqnarray}
\epsilon_t &\simeq& - 0.008 \ GeV^4, \nonumber\\
\langle{0}|{\alpha_s \over
\pi}G^a_{\mu\nu}G^a_{\mu\nu}|{0}\rangle &\simeq& 0.0283 \ GeV^4,
\end{eqnarray}
and
\begin{eqnarray}
-0.00933 &\leq \epsilon_t \leq& - 0.00724 \ (GeV^4), \nonumber\\
0.0256 \leq &\langle{0}|{\alpha_s
\over \pi}G^a_{\mu\nu}G^a_{\mu\nu}|{0}\rangle& \leq 0.0331 \ (GeV^4).
\end{eqnarray}
There exist already phenomenological estimates of the gluon
condensate [85] as well as lattice calculations of the vacuum
energy density [86] pointing out that the above mentioned standard
values (6.14) and (6.15-6-16) are too small. Our numerical predictions
are in agreement with these estimates though we think that their
numbers for the gluon condensate [85] are too big. At the same
time, it becomes quite clear why the standard values are so
relatively small, because they are due to the instanton component
of the vacuum only.
 
As it was mentioned above, a nontrivial rearrangement
of the QCD vacuum will occur if the density of the instanton-like
fluctuations achieves some critical value.
This critical value can be reached
when $\epsilon_I \simeq \epsilon_q + \epsilon_g$, i. e. when at
least
one sort of virtual quarks is presented in the true QCD vacuum.
On one hand, this assumption is supported by our numerical results
for
$ \epsilon = \epsilon_q + \epsilon_g$. On the other hand,
the numerical results for $\epsilon_I$, as given by (6.15) or
(6.16), also confirm this.
In the realistic (non-chiral) case the instanton part, along with
other contributions, may substantially (but not surely) differ
from those shown in Tables 1, 2, 3, 4 and 5.

\subsection{Summary}

 One of our main conclusions is that in QCD, unlike in QED, the
self-energy
and bound-state problems are closely related to each other. The
propagation of the constituent quark necessarily becomes a
many-body problem. Apparently, the ZME effect
correctly describes this feature of QCD at large distances and
therefore it correctly reflects the nonperturbative structure of
the QCD vacuum, at least, at first approximation.
Following Susskind and Kogut [39], we also would like to
especially emphasize the importance of quantum fluctuations of the
IR degrees of freedom and the non-abelian character of QCD.
Let us additionally underline the role of the ghost
degrees of freedom in the IR structure of the QCD vacuum.
Neglecting them "by hand", one necessarily comes to the almost
trivial ST
identities for cubic and quartic gluon vertices which
are responsible for the direct self-interaction of gluons [30].
 
  From our model it follows that the non-abelian
character of QCD alone provides quark confinement and DCSB
at large distances through the strong interaction of quarks with
the nonperturbative vacuum (enhancement of the zero modes) and at
short distances it leads to the asymptotic
freedom phenomenon because of the strong
interaction with the vacuum of the gluon degrees of freedom. This
confirms the expectation of Marciano and Pagels expressed
in their excellent review on QCD  "The gluon field, unlike the
photon,
carries color charge and has self-interactions. This distinction,
we believe, leads to confinement; but we will not see this
property using ordinary perturbation techniques"  [1].
If our QCD inspired nonperturbative model of quark confinement is
correct, at least in first approximation,
then confinement is not an additional external
requirement to complete QCD in order to be a self-consistent
theory of strong
interactions. It seems to us that QCD is a complete theory and
provides a unique description of the hadron dynamics at high
and low energies without any additional assumptions.
 
  It is well known that instantons themselves can produce
a rather complicated mechanism for DCSB in the QCD vacuum
[70, 87, 88] but they do not explain confinement.
In contrast, our model explains quark confinement and DCSB
(and many other things) on the same basis of the enhancement of
the zero modes because of the
possible nonperturbative IR divergences in the true vacuum of QCD.
From our discussion it clearly follows that our model
needs instanton-like fluctuations in the vacuum of QCD in order to
provide transitions between hadrons.
We understand the role of instantons in quite a different way [89]
than in the
RILM while, nevertheless, it seems to us that our model is able
to reconcile the instantons with the quark confinement phenomenon,
at least qualitatively.
 Generally speaking, it would be of a great
interest to find instanton-like solutions to the classical
equations, on account of the enhancement of the zero modes, if
it is possible at all, since this is essentially a quantum effect.

\section{Comparison with related models}

\subsection{The CQM}

 From a QCD theoretical field point of view, the CQM is
nothing but an approximation of the full quark propagator by
the constituent quark propagator (5.14) and the full
constituent-quark-gluon vertices
by point-like ones. The bound-state amplitudes are approximated by
the corresponding coupling constants, for example, the pion BS
amplitude (2.18-2.19) should be approximated by the pion-quark
coupling constant (5.32). Choosing a reasonable value for a
possible UV cutoff, one arrives at a calculation scheme
which
works surprisingly well (despite its simplistic structure) in the
description of static hadron properties. It is well known that
the CQM agrees with experiment within an accuracy of 20\% [90].
 
  It has been already emphasized above that the ZME
model of quark confinement and DCSB  necessarily
leads to the concept of the constituent quark. Our numerical
results clearly show that the dynamical (and consequently
constituent, too, because of the relation (8.1)) quark
really plays an important role. From the discussion in the
preceding
section about the dynamical mechanism of quark confinement and
DCSB, within the ZME effect, it follows that
the hadrons are relatively weak bound states of the
constituent quarks. A current quark in our quantum field theory
model, because of the above mentioned effect, becomes first
effective and then constituent (i.e. heavier) only at the expense
of the quantum fluctuations of the gluon fields, i.e. it
becomes constituent one without any additional admixture of more
complicated states.
Having become constituent, then its interaction with actual
gluons can be described as point-like, at least, in first
approximation.  On one hand, this
justifies the point-like character of the constituent
quark-gluon vertex in the CQM. On the other hand, it confirms
Shuryak's
observation [91] in the QCM that the coupling constant of the
constituent quark to the bare or "current" quark is equal to
one within error limits.
The constituent quark-gluon vertex
does not differ much from the point-like current quark-gluon one,
at least, in first approximation. This is shown in Fig. 20.
 
  As a side remark, recently Fritzsch [55] proposed
to interpret a constituent quark inside hadrons as
a quasi-particle which has a non-trivial internal structure on its
own, i. e. consisting of a valence quark, many quark-antiquark
pairs and of gluons - in short, it looks like one third of a
nucleon. As was it discussed above, nearly the same picture
emerges from our model which is based on the ZME effect of QCD.
The definition of the effective
quark mass, as a solution of the full quark propagator at zero
(5.10), should be appropriate for an effective mass of
Fritsch's constituent quark. This definition incorporates a
complicated dynamical structure of a constituent quark
and, in a self-consistent way, relates it to such
nonperturbative effects in QCD like quark confinement and DCSB.
 
Thus, in many aspects, our approach,
with the constituent quark playing such a significant role, looks
like as a generalization of the CQM [32, 61, 64, 90].
However our approach is directly derived from QCD. Therefore it
can
be considered as a theory giving strong arguments in favour of the
reconciliation of the CQM with QCD and thereby it puts the CQM on
a firm theoretical basis. In other words our model makes it
possible to understand how the CQM may be derived from QCD.
However, see remarks in the next section.

\subsection{The bag model}

  A nontrivial relation between our
model, on one hand, and the bag [92-94] and string [95]
models, on the other hand, would not be surprised.
In this connection some dynamical aspects of our model
should be underlined. From the above consideration it follows
that,
from a dynamical point of view, maybe the ZME effect does
not lead to string configuratons of flux tube type between quarks.
Nevertheless, there is no doubt that this dynamical process works
like a
string preventing quarks to escape from each other. It takes place
in the finite volume of the QCD vacuum but it does not
require the introduction of an explicit surface. The finiteness
of the cut-off $z_0$ results in unphysical singularities (at this
point $z_0$) of
the solutions to the quark SD equation which are due to inevitable
ghost degrees of freedom in QCD. This has nothing to do with the
bag fixed boundary. Numerically it depends on a scale at which
nonperturbative effects become essential in our model. We treat
a hadron as a dynamical process which takes place in some
finite volume of the vacuum rather than as an extended object with
an explicitly fixed surface in the vacuum. The ZME model
remains a local field theory. However, the excistence of
the vacuum energy per unit volume -- the bag constant $B$ -- is
important in our model as well. The inward positive pressure $B$
counterbalances the vacuum energy density needed for
generating the vacuum fluctuations, inspired  by the
enhancement of the zero modes in our model, i. e. the sum of the
bag constant and the nonperturbative vacuum energy density must
be zero. In agreement with the authors of paper [96], we
consider the bag constant as a universal one which
characterizes the complex nonperturbative structure of the QCD
vacuum itself and it does not depend on the hadron matter.
Moreover,
being an important characteristic of the QCD vacuum, it greatly
influences the quark-gluon plasma (QGP) equation of state [77],
bridging the gap between hadron matter and the QGP phase.
 
The bag constant is defined as the difference between the energy
density of the perturbative and the nonperturbative QCD vacuums.
We normalized the perturbative vacuum to zero, Eq. (4.8).
In our notations the bag constant becomes
\begin{equation}
B = - \epsilon_t.
\end{equation}
(Not to be mixed
with the CHPTh constant B which measures the vacuum expectation
value of the scalar densities in the chiral limit, see Subsection
C of Section 5 and Tables 1 and 2).
Our predictions are listed in Tables 3 and 4, 5
for each calculation scheme A and B, respectively. In fact, our
values for the bag constant overestimate the initial MIT bag [93]
volume energy by one order of magnitude. Nevertheless, we think
that the introduction of this quantity into physics
was a main achievement of the bag model.
 
 As in previous case, let explicitly reproduce some interesting
 particular values of the bag constant.
For a pure gluodynamics ($N_f=0$) it is:
\begin{equation}
B \simeq 0.005 \ GeV^4 \simeq (266 \ MeV)^4 \simeq 0.651 \ GeV/fm^3.
\end{equation}
 
  For the most realistic case $N_f=2$ the bag constant becomes
\begin{equation}
B \simeq 0.008 \ GeV^4 \simeq (300 \ MeV)^4 \simeq 1 \ GeV/fm^3.
\end{equation}
For simplicity's sake we reproduced its values obtained within the
calculation scheme A only. It has been noticed in [97] that
noybody
knows yet how big the bag constant moght be, but generally it is
thought it is about $ 1 \ GeV/fm^3$. The predicted value for
$N_f=2$ is in fair agreement with this expectation.

\subsection{Some other models}

 The Nambu-Iona-Lasinio model [98]
of QCD does not possess confinement phenomenon at all, so its
relation to our model remains still unclear. It is not
without sense to mention also the quark pair
condensation model of chiral symmetry breaking in QCD [99]
because in many aspects our model
is a covariant generalization and extension of that model. Our
numerical program and conclusions, concerning especially DCSB, are
almost the same as in that model which approximates the QCD
vacuum by a condensate of flavour and colour singlet
quark-antiquark pairs. The important role
of the constituent quark concept is also emphasized. The
differences between our model and theirs are: 1) Our model is
formulated
in a relativistically invariant approach to QCD at low
energies on the basis of the enhancement of zero modes due to
the self-interactions of gluons fields.. 2) We established
a close intrinsic link between DCSB and quark confinement.
In their approach
quark confinement was not considered at all. 3) As a result of our
calculation scheme, we have established that the scales of both
nonperturbative phenomena (quark confinement and DCSB) are nearly
the same indeed, while in their model these two scales
may be different.
 
 An interesting approach to nonperturbative QCD, based on a
weak-coupling treatment on the light front, was recently
proposed in Ref. 100. As claimed by the authors, it can bridge the
gap between QCD and the QCM. They also expect light-front IR
divergences to be the sources of confinement and DCSB. The
relation
between our approaches is not clear for us at this stage, neither
with the gauged nonlocal constituent quark model (GNCQM) [101].

\section{Some important problems}

   Let us begin with the important question whether the
confinement
mechanism for heavy quarks is the same as for light ones or not.
From the quark confinement theorem (Section 3) it follows that
the propagation of a single off-mass-shell heavy or light quark
is the same. However the
propagation of heavy constituent quarks inside hadrons
differs from the propagation of light constituent quarks.
First of all, the chiral limit
makes no sense for heavy quarks. This means that the quantum
fluctuations (important for the light constituent quarks) do not
influence heavy constituent quarks greatly, so the difference
(described by $\Delta$ in (8.1)) between heavy constituent and
effective quarks should become small. This is also true for the
instanton-like fluctuations which influence heavy quarks from the
loops much less than light ones (it is possible to say that, like
in the RILM [102], in our model heavy (static) quarks essentially
ignore instantons). This explains why, in general,
ground-states of mesons consisting of heavy quarks (for
example, $c \bar{c}$ systems) should have a much narrow width
than mesons consisting of light ones [54], as well as why
the potential concept should work much better for heavy
constituent quarks than for light constituent ones [44].
The spin-flavor symmetry, discovered in the heavy quark effective
theory [103], cleary shows that there might be a difference
in the dominant configurations of quantum fluctuations between
heavy constituent quarks and those of light constituent ones.
In this context let us note that recently
color-singlet glueballs were suggested (described as a loop of one
quantum of color magnetic flux) to produce the heavy quark
-confining force, even though it is difficult to distinguish
a glueball from a color and flavour singlet quark-antiquark pair
[104].
 
  Thus a difference between the confinement of heavy
and light quarks does exist but it is not principal.
We see no principal distinction in the process of
topologically nontrivial vacuum rearrangement for heavy-heavy,
light-light and heavy-light constituent quarks from which
corresponding mesons are build.
 
The quarks become constituent in the presence of each other
in our model of the QCD vacuum, i.e. they are
bound-state objects. While in the CQM
for the description of the bound-states
a linearly rising (imitating contributions of the virtual gluons)
plus a Coulomb-like (imitating contributions of the actual gluons)
potential is used along with the corresponding corrections [
44, 45]. In our approach this problem
should be considered within the BS formalism for the bound-states
(Fig. 21). So the potential concept of the QCM, based on
the point-like vertices, is too naive an approximation of the real
dynamical picture which emerges in the framework of our model.
The same conclusion, namely, that, in fact, quark interactions are
much more
complicated than just simple universal confining forces, follows
from the study of correlation functions in the QCD vacuum
[91]. Moreover we think that the CQM, based on the potential
concept, has already played its useful role
and now it should be retired from the scene like the Bohr orbits
were after the creation of the true theory of atoms - quantum
mechanics [105].
 
To investigate the bound-state problem, one should begin from the
first skeleton diagram (the  "generalized ladder
approximation", Fig. 22a) of the BS scattering kernel.
The full gluon propagator has to be replaced by (3.6-3.8),
reproducing the effect of the enhancement of the zero modes.
Quark
propagators should also be replaced by the solutions reconstructed
on the basis of this effect (Section 3) together with
the quark-gluon
vertices. These vertices should be independently taken from the
investigation of the corresponding WT identities (see our paper
[16] and  Section 3). The kernel of the BS integral
equation for mesons obtained this way, in general, will contain
the derivatives of the corresponding vertices (see, for example,
equation (3.14)). This complicated kernel will be
responsible for the interaction between the constituent quarks
in our model in the generalized ladder approximation. This has
nothing to do with the above mentioned potential (linear plus
Coloumb-like) of the QCM. A much
more complicated object appears if one takes into account the
second diagram of the skeleton expansion of the BS scattering
kernel (Fig. 22b), etc. We hope that this complication of the
interaction between constituent quarks, that emerges in our
model, will make it possible to resolve an old-standing problem of
the QCM, namely the unphysical inverse-power color analog van der
Waals potentials between separated hadrons are in
substantial contradiction with experimental data (see Ref. 106 and
references therein). The BS equation for single-hadron
states and the aspects of hadron-hadron interaction
are of great important and will be
investigated elsewhere. In our paper [19] the Goldstone nature of
the pion was confirmed by the investigation of the corresponding
axial-vector WT identity (see also Section 2).
This should be supported by the investigation of the BS
equation in the chiral limit for the pseudoscalar bound-states
in our model of confinement and DCSB, too. This topic for
further work is also important.
 
 Let us now proceed to more technical problems related to our
approach. Note that the non-chiral limit
is neither a dynamical or a technical problem for our model.
The first essential technical problem is that gauge
invariance is important for
numerical calculations of various low energy physical parameters.
The quark SD equation depends explicitly on a gauge fixing
parameter. This has two reasons. Firstly, it comes
from the full gluon propagator, secondly from the corresponding
quark-gluon vertex function. In both cases we have
shown [16] that after the completion of our renormalization
program, to remove all nonperturbative IR divergences,
the explicitly gauge-dependent terms (the
next-to-leading terms, i.e. the third line in Eq. (3.14))
disappear.
The final form of the renormalized (IR finite) quark SD equation
(3.29-3.30) does
not depend explicitly on a gauge fixing parameter.
The "gauge invariance" of the quark propagator
(more precisely of its nonperturbative IR piece) is
understood in this sense. Within the general scheme of our
approach [16], the explicit dependence on a gauge-fixing parameter
may appear
again if one takes into account contributions coming
from the ultraviolet (UV) region. This is beyond
the scope of the present calculations. In any case, nobody
knows yet how to take into account the UV piece of the quark SD
equation in a gauge-invariant way.
 
   Also, the UV renormalization program becomes trivial
in our dynamical quark propagator approach [16].
Indeed, the ground solution
(3.32-3.33) in the chiral limit at
$t_0 = \infty$ automatically approaches the free quark propagator
at infinity (asymptotic freedom) like it should.
Renormalization was effectively taken into account
by identifying the cut-off with the constant of integration of
the equation of motion as described in Section 3. So, as far
as dealing with the IR
piece of the quark SD equation (automatically satisfying
asymptotic freedom), one can completely ignore the UV
renormalization. A non-trivial UV
renormalization should be performed  when one
will take into account contributions from the UV
region in the quark SD equation, what, as mentioned above, is
beyond the
scope of this paper. Incidentally, let us stress once more
that we have shown explicitly that the numerical values of basic
chiral QCD parameters are determined mainly by the contributions
coming from the IR region, while the contributions from
short distances (for which UV renormalization is
essential) can only be treated as small perturbative corrections.
 
 Our treatment of the strong IR singularity (3.1), in the
sense of distribution theory, within
the system of equations in the quark sector, was carried out in
Euclidean space. The corresponding expansion (3.8) takes place
in this space since in Minkovskian domain it becomes much
more complicated and may contain, for example, derivatives of
the $\delta$- function [43]. This means that the final system of
equations will be completely different from that obtained in
Euclidean space and, therefore, requires separate
consideration. However there may be a vital problem in the
continuation of the obtained Euclidean
solutions of the quark SD equation to the time-like region (Wick
rotation). As it was pointed out in our paper [16], the analytical
continuation of the nonperturbative solutions (with a typical
exponential non-analiticity in the coupling constant (3.32-3.33))
could not be performed without additional conditions.
For example, our solutions for the IR finite quark propagator
(3.32-3.33) can, in principle, be analytically continued to the
time-like region ($t \rightarrow - \infty$) if one changes
simultaneously the sign of the ghost self-energy at zero point
in (3.31), too. This can be done formally because the SD equation
for the
ghost self-energy at zero point has two solutions: a  positive and
a negative one [16]. This once more emphasizes the role of the
ghost fields in the IR
nonperturbative structure of the QCD ground state - i.e. true
vacuum.
However, as it turns out, we have no problems with the Wick-
rotation
within our low energies QCD model. In order to calculate
physical quantities we always perform an integration over the
nonperturbative region whose finite size is determined by the
confinement scale $\Lambda_c$ (Section 6). Ending the discussion
of these important problems, let us ask a, perhaps, heretic, but
relevant question. Does it make sense
at all to worry about Wick rotation of the solutions
for the unobservable quarks and gluons?
The numerical values of observable characteristics of hadrons (for
example, masses, decay constants, etc), calculated from first
principles, do not depend, of course, on the metrics used.
 
 There are many problems to be solved and questions
to be answered in our model of quark confinement and DCSB.
It is quite possible that our interpretation of the enhancement
of the
zero modes (3.1) or (3.8) as the enhancement of the contributions
to the quark self-energy, due to self-interaction of gluon fields
is not completely correct, but, we think, it works in the
right
direction. By all means, we believe that our model (even
without the proposed interpretation ) is a first step in that
direction.

\acknowledgments

  The authors would like to thank J. Zim\'anyi, K. Szeg\H o,
Gy. P\'ocsik, K. Lad\'anyi, B. Luk\'acs, K. T¢th, T. Dolinszky,
G. P\'alla, J. R\'evai, T. Bir\'o
and other members of the Theoretical Department
of the RMKI for their constant interest, valuable discussions
and support. One of the authors (V. G.) is also grateful to
V.A. Rubakov, N. Brambilla, W. Lucha,
F. Schoberl for useful discussions and remarks and especially
to N.B. Krasnikov for detailed discussion of the QCD true vacuum
properties and his persistent advice to take seriously relation
(4.25) despite its perturbative origin. This work was supported
by the Hungarian Science Fund (OTKA) under Grants No. T 16743,
No. T 1823, and No. T 016206.

\vfill
\eject

\vfill
\eject

\begin{table}
\caption{Calculation scheme A}
\begin{tabular}{|l|l|c|r|r|} \hline
$F^o_{\pi}$ & 88.3 & 92.42 & 93.3 & MeV \\ \hline
 
$\Lambda_{CSBq}$ & 724.274 & 758.067 & 765.284 & MeV \\
 
$m_d$ & 362.137 & 379.0335 & 382.642 & MeV \\
 
${\langle \overline qq \rangle}_0$ & $(-208.56)^3$ & $(-218.29)^3$
& $(-220.36)^3$ & $MeV^3$ \\
 
$\epsilon_q$ & $-0.0012$ & $-0.00143$ & $-0.0015$ & $GeV^4$ \\
 
$\epsilon_g$ & $-0.0013$ & $-0.00157$ & $-0.0016$ & $GeV^4$  \\
 
$\epsilon$ & $-0.0025$ & $-0.0030$ & $-0.0031$ & $GeV^4$  \\
 
$\langle{0}|{\alpha_s \over
\pi}G^a_{\mu\nu}G^a_{\mu\nu}|{0}\rangle$ & 0.009 & 0.0106 &
0.011 & $GeV^4$  \\
 
B & 1163.51 & 1217.78 & 1229.23 & MeV \\
 
$m^0_u$ & 6.65 & 6.36 & 6.30 & MeV \\
 
$m^0_d$ & 10.08 & 9.63 & 9.54 & MeV \\
 
$m^0_s$ & 202.85 & 193.75 & 191.94 & MeV \\
 
$g^o_{{\pi}qq}$ & 3.0177 & 3.1584 & 3.1885 & $MeV^0$ \\
 
G & $2.05 \times 10^5$ & $2.25 \times 10^5$  & $2.29 \times 10^5$
& $MeV^2$ \\ \hline
\end{tabular}
\end{table}

\vfill
\eject

\begin{table}
\caption{Calculation scheme B}
\begin{tabular}{|l|c|r|} \hline
$F^o_{\pi}$ = 88.3 & $F^o_{\pi}$ =  92.42 & $F^o_{\pi}$ = 93.3
\\ \hline
 
$707 \le \Lambda_c \le 742.68$ & $737.9 \le \Lambda_c \le
768.4$  & $744.4 \le \Lambda_c \le 773.86$  \\
 
$328.62 \le m_d \le 400$ & $340 \le m_d \le 400$ & $342.416
\le m_d \le 400$ \\
 
$(-210.34)^3 \le {\langle \overline qq \rangle}_0 \le (-206.9)^3$
&$(-219.3)^3 \le {\langle \overline qq \rangle}_0 \le (-216.34)^3$
&$(-221.2)^3 \le {\langle \overline qq \rangle}_0 \le (-218.33)^3$
\\
 
$-0.00135 \le \epsilon_q \le -0.00096$ & $-0.0016 \le \epsilon_q
\le -0.00128$ & $-0.0017 \le \epsilon_q \le -0.00136$ \\
 
$-0.0024 \le \epsilon_g \le -0.00045$ & $-0.00226 \le \epsilon_g
\le -0.00044$ & $-0.00221 \le \epsilon_g \le -0.000437$  \\
 
$-0.00336 \le \epsilon \le -0.0018$ & $-0.00354 \le \epsilon \le
-0.002$ & $-0.00356 \le \epsilon \le -0.0021$ \\
 
$0.0064 \le {\langle{0}|G^2|{0}\rangle} \le 0.0128$
& $0.007 \le {\langle{0}|G^2|{0}\rangle} \le 0.0192$
& $0.00746 \le {\langle{0}|G^2|{0}\rangle} \le 0.0199$ \\
 
$1135.95 \le B \le 1193.56$ & $1185.44 \le B \le 1234.76$  &
$1195.57 \le B \le 1243.34$ \\
 
$6.48 \le m^0_u \le 6.81$ & $6.27 \le m^0_u \le 6.53$  &
$6.22 \le m^0_u \le 6.47$ \\
 
$9.83 \le m^0_d \le 10.33$ & $9.5 \le m^0_d \le 9.89$ &
$9.43 \le m^0_d \le 9.81$ \\
 
$197.67 \le m^0_s \le 207.7$ & $191 \le m^0_s \le 199$ &
$189.76 \le m^0_s \le 197.34$ \\
 
$2.74 \le g^o_{{\pi}qq} \le 3.33$ & $2.83 \le g^o_{{\pi}qq} \le
3.33$ & $2.89 \le g^o_{{\pi}qq} \le 3.33$ \\
 
$2.0 \times 10^5 \le G \le 2.1 \times 10^5$ &
$2.19 \times 10^5 \le G \le 2.28 \times 10^5$ &
$2.23 \times 10^5 \le G \le 2.32 \times 10^5 $ \\ \hline
\end{tabular}
\end{table}

\vfill
\eject

\begin{table}
\caption{Calculation scheme A. Predictions}
\begin{tabular}{|l|l|r|r|} \hline
$F^o_{\pi}$ & 92.42 & 93.3 & MeV \\ \hline
 
$\epsilon_t = \epsilon_I + \epsilon_g + N_f \epsilon_q$ &
$-0.00497 - N_f 0.00143$ & $-0.005 - N_f 0.0015$ & $GeV^4$ \\
 
$\langle{0}|{\alpha_s \over
\pi}G^a_{\mu\nu}G^a_{\mu\nu}|{0}\rangle$ & $0.01767 + N_f 0.00508$ &
$0.01777 + N_f 0.00533$ & $GeV^4$  \\
 
B  & $0.00497 + N_f 0.00143$ & $0.005 + N_f 0.0015$ & $GeV^4$  \\ \hline
\end{tabular}
\end{table}

\vfill
\eject

\begin{table}
\caption{Calculation scheme B. Predictions}
\begin{displaymath}
\begin{array}{|c|} \hline
\; F^o_{\pi} = 92.42 \; \\ \hline
\; -0.00566 - N_f 0.00128 \le \epsilon_t
\le -0.00384 - N_f 0.0016 \;
 \\
\; 0.0136 + N_f 0.00568 \le {\langle{0}|G^2|{0}\rangle} \le 0.02
+ N_f 0.0045 \;
\\
\; 0.00384 + N_f 0.0016 \le B \le 0.00566 + N_f ).00128 \;
\\ \hline
\end{array} \nonumber
\end{displaymath}
\end{table}

\vfill
\eject

\begin{table}
\caption{Calculation scheme B. Predictions}
\begin{displaymath}
\begin{array}{|c|} \hline
\; F^o_{\pi} = 93.3 \; \\ \hline
\; -0.00661 - N_f 0.00136 \le \epsilon_t
\le -0.003837 - N_f 0.0017 \;
 \\
\; 0.0136 + N_f 0.006 \le {\langle{0}|G^2|{0}\rangle} \le 0.0235
+ N_f 0.0048 \;
\\
\; 0.003837 + N_f 0.0017 \le B \le 0.00661 + N_f ).00136 \;
\\ \hline
\end{array} \nonumber
\end{displaymath}
\end{table}
 
 \vfill
 \eject

\begin{figure}

\caption{ The SD equation for the quark propagator in momentum
          space.}
 
\bigskip
 
\caption{The exact expression for the pion decay constant
$F_{\pi}$ in the current algebra (CA) representation.
Here $G^j_5$, $S$ and $J^i_{5\mu}$ are the pion-quark bound-state wave
function, the quark propagator and the  axial-vector current, respectively.
The slash denotes differentiation with respect
to momentum $q_{\nu}$  and setting $q = 0$.}

\bigskip
 
\caption{The pion decay constant $F_{CA}$ as a function of
$k_0$, drawn only for the most reasonable region, selected
first for the dynamically generated quark masses (6.17).
The interval (6.18) is also explicitly shown.}

\bigskip
 
\caption{Quark condensate ${\langle \overline qq
 \rangle}_0^{1/3}$ as a function of  $k_0$,
drawn only for the most reasonable region, selected
first for the dynamically generated quark masses (6.17).}
 
\bigskip
 
\caption{The vacuum energy density, due to confining quarks
with dynamically generated masses, $\epsilon_q$ as a function of
$k_0$, drawn only for the most reasonable region, selected
first for the dynamically generated quark masses (6.17).}

\bigskip
 
\caption{The vacuum energy density of the nonperturbative
gluons $\epsilon_g$ as a function of $k_0$,
drawn only for the most reasonable region, selected
first for the dynamically generated quark masses (6.17).
The interval (6.18) is also explicitly shown.}

\bigskip
 
\caption{The vacuum energy density $\epsilon$
as a function of $k_0$,
drawn only for the most reasonable region, selected
first for the dynamically generated quark masses (6.17).}

\bigskip
 
\caption{The vacuum energy density, due to the nonperturbative
gluons contribution, $\epsilon_g$ as a function of $z_0$.}

\bigskip
 
\caption{The vacuum energy density, due to the nonperturbative
gluons
contribution, $\epsilon_g$ as a function of $k_0$. $\Lambda_c$ is
the confinement scale. A newly obtained interval for $m_d$ (6.22)
is also
shown. A similar figure can be drawn for the case when the pion
decay constant is approximated by the experimental value. }

\bigskip
 
\caption{The pion decay constant $F_{CA}$
as a function of $k_0$. $\Lambda_c$ is
the confinement scale. A newly obtained interval for $m_d$ (6.22)
is also
shown. A similar figure can be drawn for the case when the pion
decay constant is approximated by the experimental value. }

\bigskip

\caption{The simplest contribution to the quark
self-energy induced by a virtual (internal) gluon. }
 
\bigskip
 
\caption{The simplest contributions to the quark
self-energy induced by the self-interactions of virtual (internal)
  gluons. }
 
\bigskip
 
\caption{The simplest contributions to the quark
self-energy induced by an actual (external) gluons emitted by
another quark.}

\bigskip
 
\caption{The simplest contributions to the quark
self-energy due to the processes of the convertation
(transformation) of the virtual gluons into the actual ones and
vice versa.}
 
\bigskip
 
\caption{ The quark diagram for the decay of the $J/ \Psi$ meson.
The dashed line schematically shows the $c \bar{c}$ pair emerged
from the nonperturbative QCD vacuum. The true number of
intermediate gluons is dictated by colour and charge
conservation.}
 
\bigskip
 
\caption{ The quark diagram for the decay of the $\Psi''$ meson.
The dashed line schematically shows the $u \bar{u}$ pair emerged
from the nonperturbative QCD vacuum. }
 
\bigskip
 
\caption{ The quark diagram for the decay of the pseudoscalar (P)
particle. The arrows show the reciprocal orientation of the
spins. The dashed line schematically shows the pair emerged
from the QCD vacuum. The true number of intermediate
gluons is dictated by colour and charge conservation.}
 
\bigskip
 
\caption{ The quark diagram for the decay of the vector (V) meson.
 The arrows show the reciprocal orientation of the spins.
The dashed lines schematically show the pairs emerged
from the QCD vacuum. The true number of intermediate
gluons is dictated by colour and charge conservation.
The intermediate vector state is inevitable.}

\bigskip
 
\caption{Two-particle irreducible (2PI) quark, ghost and
gluon vacuum diagrams. }
 
\bigskip
 
\caption{The interaction of a constituent quark (thick line)
and current quark (thin line) with actual gluon.}
 
\bigskip
 
\caption{The Bethe-Salpeter (BS) integral equation for flavored
mesons.}

\bigskip
 
\caption{The BS scattering kernel.}
 
\end{figure}


\begin{references}
\bibitem{48}
Apparently, the point $t_0 = \Lambda$ is a stationary point of
the effective potential (4.1), but to establish its nature
(whether it is a saddle point or local extremum) is beyond the
scope of the present paper. In any case, we know that $\Lambda$
should be identified with $t_0$ in order to avoid the influence of
unphysical singularities (due to the necessary ghost degrees
of freedom) on the elements of the $S$-matrix, i.e. their
contributions to the numerical values of the physical quantities.
In our previous publication (Ref. 16) this point was
prematurely assumed to be a local minimum. For the analytical
investigation of this problem (i.e. the nature of the
stationary point), in various forms of the effective
potential, see the comprehensive review of Haymaker in Ref. 6.
\bibitem{89}
   It is possible to say that we treat the instanton component of
the QCD vacuum not as a "liquid" but rather as a "forest" in which
instantons and anti-instantons are considered as "trees" with
$SU_c(3)$ orientation, position and scale size.
\end{references}
\end{document}